\documentclass[manuscript,screen]{acmart}

\usepackage[most]{tcolorbox}

\newcommand{\hl}[1]{\underline{\textcolor{red}{#1}}} 
\definecolor{darkgreen}{rgb}{0.0, 0.5, 0.0}

\newcommand{\rqone}{Can we build a benchmark for automated SATD repayment that is more representative and less noisy?}
\newcommand{\rqtwo}{Can we design evaluation metrics that are both effective and interpretable for automated SATD repayment?}
\newcommand{\rqthree}{Does a cleaner dataset lead to a better performance in automated SATD repayment?}
\newcommand{\rqfour}{How do fine-tuned small models compare to prompt-based approaches using large LLMs for automated SATD repayment?}

\AtBeginDocument{%
  }

\setcopyright{acmlicensed}
\copyrightyear{2025}
\acmYear{2025}
\acmDOI{XXXXXXX.XXXXXXX}

\acmConference[]{Preprint submitted to ACM Transactions on Software Engineering and Methodology (TOSEM)}{Jan 16, 2025}{}
\acmISBN{978-1-4503-XXXX-X/18/06}

\begin{document}

\title{Understanding the Effectiveness of LLMs in Automated Self-Admitted Technical Debt Repayment}

\author{Mohammad Sadegh Sheikhaei}
\email{sadegh.sheikhaei@queensu.ca}
\orcid{0000-0003-4219-642X}
\author{Yuan Tian}
\email{y.tian@queensu.ca}
\affiliation{%
  \institution{School of Computing, Queen’s University}
  \city{Kingston}
  \state{ON}
  \country{Canada}
}

\author{Shaowei Wang}
\affiliation{%
  \institution{Department of Computer Science, University of Manitoba}
  \city{Winnipeg}
  \state{MB}
  \country{Canada}}
\email{Shaowei.Wang@umanitoba.ca}

\author{Bowen Xu}
\affiliation{%
  \institution{Department of Computer Science, North Carolina State University}
  \city{Raleigh}
  \state{NC}
  \country{USA}
}
\email{bxu22@ncsu.edu}


\begin{abstract}
Self-Admitted Technical Debt (SATD), cases where developers intentionally acknowledge suboptimal solutions in code through comments, poses a significant challenge to software maintainability. Left unresolved, SATD can degrade code quality, increase maintenance costs, and hinder long-term software sustainability. Automating SATD repayment is therefore essential to alleviate developer burden. While Large Language Models (LLMs) have shown promise in tasks like code generation and program repair, their potential in automated SATD repayment remains underexplored.

In this paper, we identify three key technical challenges in training and evaluating LLMs for SATD repayment: (1) dataset representativeness and scalability, (2) removal of irrelevant SATD repayment samples, and (3) limitations of existing evaluation metrics, i.e., BLEU and CrystalBLEU. To address the first two dataset-related challenges, we adopt a language-independent SATD tracing tool and design a 10-step filtering pipeline to extract SATD repayments from repository commit histories, leveraging LLM-as-judge for relevance filtering. This results in two new large-scale SATD repayment datasets: 58,722 items for Python and 97,347 items for Java. To improve evaluation, we introduce two diff-based metrics, BLEU-diff and CrystalBLEU-diff, which measure code changes rather than whole code, thereby mitigating the impact of SATD-irrelevant source code. Additionally, we propose a new metric, Line-Level Exact Match on Diff (LEMOD), which is both interpretable and informative, providing fine-grained insights into SATD repayment quality. Using our new benchmarks and evaluation metrics, we evaluate two types of automated SATD repayment methods: fine-tuning smaller models (a few hundred million parameters) and prompt engineering with large-scale models, including four SOTA open-source LLMs and GPT-4o-mini. Our results reveal that while fine-tuned small models achieve Exact Match (EM) scores comparable to prompt-based approaches, they underperform on BLEU-based metrics and LEMOD. The best EM performance is achieved by Gemma-2-9B, correctly addressing 10.1\% of Python SATDs and 8.1\% of Java SATDs with a simple prompt. The only previous study in this domain focused solely on Java, achieving an EM score of 2.3\%. When using BLEU-diff, CrystalBLEU-diff, and LEMOD metrics, Llama-3.1-70B-Instruct and GPT-4o-mini deliver the highest performance. While our three proposed metrics demonstrate strong correlations with EM (0.65\textasciitilde0.84), both BLEU and CrystalBLEU show almost no correlation with this intuitive metric (0.01\textasciitilde0.08), highlighting their largely unpredictable behavior when applied directly to the entire code rather than the diff. Our work contributes a robust benchmark, improved evaluation metrics, and a comprehensive evaluation of LLMs, advancing research on automated SATD repayment.

\end{abstract}

\ccsdesc[500]{Software and its engineering~Software development process management}
\ccsdesc[500]{Computing methodologies~Artificial intelligence}

\keywords{Self-Admitted Technical Debt, Large Language Models, SATD Repayment Dataset, SATD Repayment Evaluation}


\maketitle

\section{Introduction}\label{sec:introduction}
Technical Debt (TD) refers to the intentional use of suboptimal solutions in software design or coding, typically to meet tight deadlines or address immediate resource constraints~\cite{Cunningham-1992}. Similarly to financial debt, technical debt accumulates ``interest'' in the form of increased future costs due to earlier poor design and implementation choices~\cite{Buschmann-2011, Brown-2020}. The awareness of the technical debts in a software system is often achieved by developers self-admitting them in software artifacts, such as code comments (e.g., ``\textit{TODO: Optimize it}''), commit messages, and issue reports. This type of TD has been referred to as \textit{Self-Admitted Technical Debt (SATD)}~\cite{Potdar-2014}. This notion enables many data-driven studies that empirically analyze the introduction and removal of technical debts by tracing the lifespan of SATDs within the history of a software project. Prior research finds that SATD is present in 2.4\% to 31\% of the source files~\cite{Potdar-2014}, with a median lifespan ranging from 18 to 172 days~\cite{maldonado2017empirical}. Prolonged unresolved SATD negatively impacts code quality and maintainability, increasing future technical costs and engineering effort~\cite{Li-2023-estimating-effort, Chowdhury-2024}.

These data-driven empirical studies have not only shed light on the statistics and lifespan of SATD, but also motivate an important question: \textit{Can we leverage historical or cross-project technical debt repayments to automatically address SATDs in a timely manner?} This question becomes especially pertinent with the advent of Large Language Models (LLMs), which have demonstrated significant success in automating various code-related tasks, such as code generation, program repair, and code refactoring. However, SATD repayment introduces unique challenges that distinguish it from these tasks. SATD encompasses diverse categories, including requirement debt and design debt, which are broader and more abstract than typical code-related issues. Furthermore, SATD comments often describe best practices, design conventions, or non-functional concerns, rather than providing explicit functional summaries or bug descriptions. As a result, SATD repayment demands specialized approaches that extend beyond conventional methods for program repair and refactoring.

To train and evaluate models for automated SATD repayment, it is crucial to use datasets derived from real-world scenarios.\footnote{Theoretically, LLMs could be used to generate synthetic SATD samples; however, these samples may fail to capture the diversity and complexity of real-world SATDs and are likely to reflect the inherent biases present in the LLMs' training data~\cite{long2024llms}.} Recently, Mastropaolo et al.~\cite{Mastropaolo-2023} addressed this need by collecting a dataset of 5,039 SATD instances extracted from 595 popular Java projects hosted on GitHub. Their dataset was constructed in two major steps: (1) identifying SATDs in the repository commit history and mining code changes involving the removal of SATD comments, and (2) applying heuristics to determine cases where the SATD was addressed. Using this dataset, they compared smaller fine-tuned LLMs (e.g., CodeT5-base with 220 million parameters) with GPT-3.5 in a zero-shot setting (without continual training on task-specific data). They reported that fine-tuned smaller LLMs outperformed GPT-3.5. However, even the best-performing model, the CodeT5 fine-tuned on 3,537 SATD repayment samples, successfully addressed only 2.24\% of SATDs in their evaluation set. Based on these results, Mastropaolo et al. concluded that automated SATD repayment remains a highly challenging task.

While Mastropaolo et al.’s work represents a pioneering contribution by introducing a benchmarking system for evaluating LLMs in SATD repayment, we argue that their conclusion may be premature. A more rigorous evaluation of LLM performance is necessary, as we identify three technical challenges not addressed in their data creation and benchmarking process. These challenges hinder a more realistic assessment of the capabilities of LLMs in automated SATD repayment.

\begingroup
\renewcommand{\theenumi}{Ch\#\arabic{enumi}}
\renewcommand{\labelenumi}{(\theenumi)}

\begin{enumerate}
    \item \textbf{Lack of representativeness due to limitations of the SATD tracing tool.} Mastropaolo et al. leveraged a tool named SATDBailiff~\cite{AlOmar-2022} to extract code changes that contain the removal of SATDs from repositories. However, SATDBailiff relies on a Java parser, which requires significant computational time to process large repositories. As a result, in Mastropaolo’s experiment, the tool failed to process 6,162 out of 6,971 targeted repositories (88\%) within the allocated time limit of one hour per repository. The low repository-level extraction success rate of 12\% may result in a biased dataset, potentially underrepresenting more complex repositories. Furthermore, since SATDBailiff is language-dependent, the analysis was restricted to Java projects, excluding projects written in other programming languages, such as Python, which are widely used and could provide valuable insights into SATD repayment.
    
    \item \textbf{Inclusion of Non-SATD repayments due to overly permissive heuristics.} Mastropaolo et al. identified SATD repayment cases by applying a heuristic that checks whether an SATD comment is removed and an update is present in the corresponding method in the subsequent version. However, this heuristic is overly permissive and often introduces false positives, where the code containing the SATD comment has been updated, but the changes are unrelated to addressing the SATD. Figure~\ref{fig:code-update-unrelated-to-SATD-repayment} illustrates an example where the deleted TODO comment suggested returning the error messages; however, the updated code did not produce any error messages, indicating that the SATD was not addressed or repaid. To assess the severity of this issue, we replicated Mastropaolo et al.'s filtering steps and manually analyzed 100 SATD removal samples (ref. Section~\ref{subsec:RQ1}). In the Java dataset, we observed that 38 samples were unclear or unrelated to SATD repayment. When we applied this heuristic to a newly created Python SATD dataset, the accuracy declined further, with only 45 out of 100 mined SATD removals corresponding to actual repayments.

    \begin{figure}[h]
      \centering
      \includegraphics[width=0.8\linewidth]{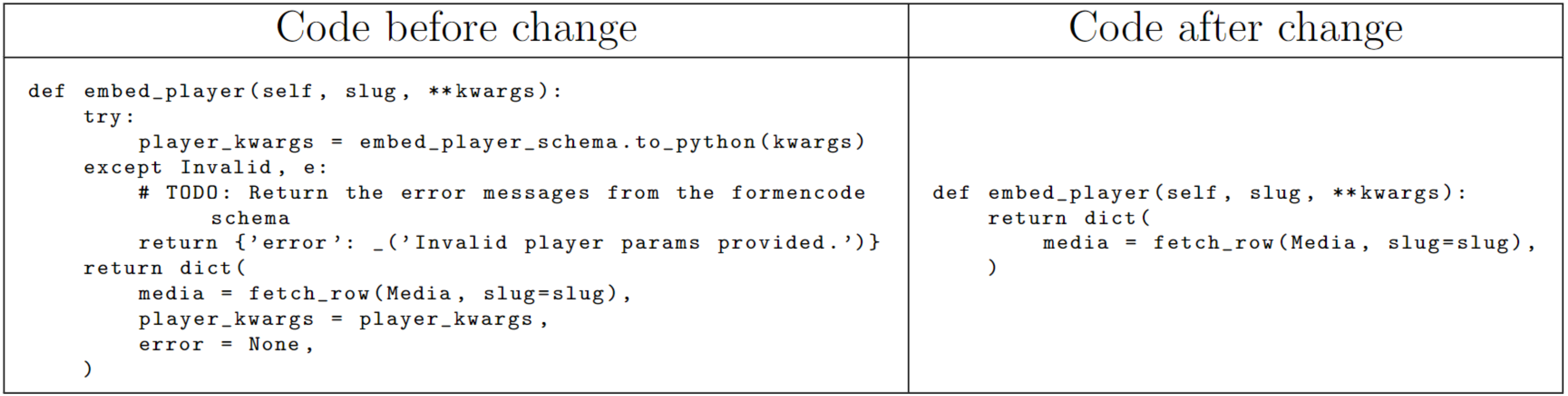}
      \caption{Example of a Java SATD removal where code update is unrelated to SATD repayment}
      \label{fig:code-update-unrelated-to-SATD-repayment}
      \Description{}
    \end{figure}
    
    \item \textbf{Inadequacy of evaluation metrics for measuring code similarity to ground truth.} Mastropaolo et al. employed two evaluation metrics: Exact Match (EM) and CrystalBLEU. The Exact Match metric is simple and intuitive, assigning a score of 1 if the generated code perfectly matches the ground truth (method containing the SATD with it being addressed) and 0 if there is any discrepancy, no matter how minor (e.g., a single character difference). While straightforward, EM often underestimates model performance because there can be multiple valid solutions and coding styles for addressing SATD. Although preprocessing steps, such as removing whitespace and comments, can reduce false mismatches, small variations in code structure or naming conventions can still result in non-matches. Although CrystalBLEU~\cite{Eghbali-2023} is designed to measure fine-grained similarity between generated code and the ground truth, Mastropaolo et al. applied it directly to the entire source code (after SATD repayment) and the ground truth. This \textit{whole-code based evaluation approach} introduces potential bias because portions of the method that are unrelated to the SATD can disproportionately influence the similarity score, making the results difficult to interpret. For instance, we observed that ignoring the SATD comment entirely, i.e., comparing the source code in the method before SATD repayment with ground truth, often results in a higher CrystalBLEU score compared to cases where the SATD is being addressed (ref. Section~\ref{subsec:RQ2}). 
    
\end{enumerate}
\endgroup

To systematically address the three identified technical challenges, we investigate the following four research questions:

\begin{itemize}
\item RQ1: \rqone
\item RQ2: \rqtwo
\item RQ3: \rqthree
\item RQ4: \rqfour
\end{itemize}

To answer RQ1, we propose a new large-scale benchmark dataset specifically tailored for SATD repayment, addressing the first two technical challenges. Instead of using SATDBailiff (\textbf{Ch\#1}), we leverage the SATD Tracker~\cite{Sheikhaei-2023} for SATD repayment candidate mining, which offers two significant advantages. First, SATD Tracker uses a language-independent approach, enabling the generation of SATD datasets across multiple programming languages, including Java and Python. Second, it is significantly faster, allowing efficient processing of large-scale datasets. Using the same repository selection criteria as Mastropaolo et al., SATD Tracker successfully processed 97.5\% of 13,748 Python repositories and 96.4\% of 7,694 Java repositories in just a few days, far outperforming the 50 days required by SATDBailiff. This improved scalability resulted in a substantially larger dataset: 1,847,392 SATD removals for Java (compared to 139,804 previously detected) and 1,059,299 removals for Python. To ensure dataset quality and remove non-SATD repayments (\textbf{Ch\#2}), we applied LLMs with zero-shot prompting to filter out code updates unrelated to SATD repayment. This filtering approach achieved F1 scores of 86.2\% for Java and 84.8\% for Python in detecting SATD repayments, resulting in a cleaner dataset.

To answer RQ2 and address the limitations of existing evaluation metrics (\textbf{Ch\#3}), we propose a diff-based evaluation approach. Specifically, we compute the differences (diffs) between the code before repayment and the ground truth, as well as between the code before repayment and the LLM-generated code. BLEU or CrystalBLEU scores are then calculated on these diffs, allowing us to focus solely on updated lines while ignoring unchanged code. Our experiments show that this method produces more interpretable results and correlates better with EM scores, i.e., achieving correlations of 0.65 to 0.83 for diff-based evaluation, compared to less than 0.1 for prior whole-code based evaluation. Moreover, we introduce Line-Level Exact Match on Diff (LEMOD), a new metric that combines the intuitiveness of EM with the granularity of diff-based evaluation. LEMOD computes precision, recall, and F1 scores (well-established metrics in machine learning) at the line level, providing a clear and meaningful assessment. Notably, LEMOD correlates strongly with EM, achieving correlation values of 0.837 for Python and 0.748 for Java.

With a cleaner dataset and improved evaluation metrics, we revisit the benchmarking of automated SATD repayment by answering RQ3 and RQ4. Mastropaolo et al.~\cite{Mastropaolo-2023} reported that their fine-tuned model achieved an EM of only 2.3\%. Given our improved dataset, we hypothesize that models can achieve better performance. Indeed, our experiments in RQ3 show that GPT-4o-mini achieves an EM of 7.5\% on the new benchmark (using the filtered Mastropaolo's dataset), significantly outperforming its performance on their original dataset (4.6\%).  Additionally, the fine-tuned model trained on the filtered dataset achieves an EM score of 6.2\%, surpassing its performance when fine-tuned on a controlled subset of the original dataset (4.9\%).

While RQ3 utilizes the Mastropaolo (Java specific) dataset to study the impact of data cleaning on the performance of automated SATD repayment, RQ4 employs our extensive Python and Java datasets to evaluate the effectiveness of LLMs in automated SATD repayment.
Specifically, we compare two types of approaches for automated SATD repayment: fine-tuning smaller models and prompt engineering with larger models. For fine-tuning, we employ two models from the CodeT5p family (an updated version of CodeT5), while for prompt engineering, we consider five SOTA LLMs, including both open-source and closed-source ones. Our experimental results reveal that while fine-tuned small models (e.g., from the FlanT5p family) achieve competitive EM scores, they underperform on more detailed metrics such as BLEU-diff, CrystalBLEU-diff, and LEMOD when compared to prompt-based larger LLMs. This finding aligns with some previous studies~\cite{liu2023empirical} but contradicts others~\cite{Mastropaolo-2023,Sheikhaei-2024}, including the conclusion drawn by Mastropaolo et al. 

Our work makes the following key contributions:
\begin{itemize}
    \item We created large, clean datasets of SATD repayment by analyzing all commits in the master branch of 14,097 Python and 7,983 Java repositories. Initially, we identified 1,607,408 SATD items in Python and 2,672,485 in Java projects. Following our 10-step filtering process, the final Python dataset includes 58,722 SATD repayment samples, and the Java dataset contains 97,347 samples, about 20 times larger than the Mastropaolo's dataset.
    \item We proposed BLEU-diff and CrystalBLEU-diff evaluation metrics, which are more reliable and interpretable than applying BLEU or CrystalBLEU to the whole code. We also introduced a new metric, Line-Level Exact Match on Diff (LEMOD), which provides more detailed information than the other metrics studied.
    \item We leveraged several open-source LLMs and a proprietary model (GPT4o-mini) with various prompts to examine the effectiveness of large language models in automated SATD repayment, and compared these results with those from smaller, fine-tuned models. The best Exact Match (EM) performance was achieved by Gemma-2-9B, which correctly addressed 10.1\% of Python SATDs and 8.1\% of Java SATDs using a simple prompt. For BLEU-diff, CrystalBLEU-diff, and LEMOD, Llama-3.1-70B-Instruct and GPT-4o-mini deliver the highest performance.  
    \item The datasets, results, and source code associated with this study are available at GitHub~\footnote{\href{https://github.com/RISElabQueens/SATD-Auto-Repayment}{https://github.com/RISElabQueens/SATD-Auto-Repayment}}.
\end{itemize}

The remainder of this paper is organized as follows: Section~\ref{sec:related-work} provides the background and related work. Our research questions and the approach for each are presented in Section~\ref{sec:research-questions}. Section~\ref{sec:study-setup} describes the study setup, including datasets, evaluation metrics, and selected LLMs. The results and analysis are presented in Section~\ref{sec:results}. In Section~\ref{sec:discussion}, we discuss various topics, including the oracle prompt template and SATD repayment coverage across all experiments. Section~\ref{sec:threats} presents the threats to the validity of our study. Finally, Section~\ref{sec:conclusion} concludes the paper and outlines future work.

\section{Background and Related Work}\label{sec:related-work}
\subsection{Empirical Study on SATD Removal and Automated SATD Repayment}
SATD repayment is critical for software developers, as incremental SATDs impact code maintainability and long-term software quality. Empirical studies on SATD removal reveal the complexities and developer behaviors around SATD management. Maldonado et al.\cite{maldonado2017empirical} found that most SATDs in open-source projects are eventually removed, often by the original introducers, indicating developers’ active management of technical debt. Zampetti et al.\cite{Zampetti-2018} further discovered that SATD removals frequently happen as part of unrelated code changes, requiring varied modifications that range from simple condition updates to extensive control-flow or API adjustments, emphasizing the challenge in automating SATD repayment. Contrary to the notion of “accidental” removals, Tan et al.\cite{tan2021practitioners} and Pina et al.\cite{pina2022technical} reported that developers tend to make deliberate decisions on SATD repayment, addressing debt as soon as possible. Studies on SATD in specific domains also offer insights: Liu et al.\cite{liu2021AnES} noted that requirement debt in deep learning frameworks is most commonly removed, while Yasmin et al.\cite{yasmin2022first} highlighted time savings potential by recognizing duplicate SATD comments. 

Manual removal of SATDs is often inconsistent and labor-intensive, making automation a promising approach to enable developers to address SATDs more efficiently and at earlier stages. In the literature, three notable studies propose automated methods for SATD repayment. The first study, by Zampetti et al. \cite{Zampetti-2020}, presents a model that classifies SATD repayment types into six categories based on the SATD comment and its associated source code. This model provides recommendations on the type of changes needed, offering structured guidance for repayment; however, it stops short of generating the actual code modifications required to implement these changes.

Mastropaolo et al. \cite{Mastropaolo-2023} move beyond categorization by automatically recommending specific code fixes using pre-trained models. They employ a SATD tracking tool, SATDBailiff, to identify deleted SATD comments and apply heuristic rules to determine addressed (repaid) instances. They evaluate both CodeT5 and ChatGPT (3.5) for their capability to generate the necessary code modifications directly. Mastropaolo et al. evaluate the performance of SATD repayment models using two primary metrics: Exact Match (EM) and CrystalBLEU \cite{Eghbali-2023} scores. EM assesses the percentage of instances where the model’s output precisely matches the ground truth code change, while CrystalBLEU evaluates the similarity of generated and target code by rewarding n-gram overlaps while reducing biases from trivial matches. They reported that CodeT5, fine-tuned on generic code changes (284,190 code changes) and SATD-specific data (5,039 identified repaid SATD), achieves a limited success rate (2\%-8\% in terms of EM) in automated SATD repayment, depending on model configurations. ChatGPT, used in a zero-shot setting, performs significantly worse, underlining the challenges of SATD repayment for general-purpose models. 

More recently, OBrien et al. \cite{obrien2024prompt} examined the potential of TODO comments as prompts to guide automated code completion tools, specifically GitHub Copilot, for SATD repayment. Their dataset includes 36,381 TODO comments extracted from Python repositories, with a manually curated sample of 380 comments to evaluate Copilot's response to SATD-related prompts. Through experiments, they assess whether TODO comments, when included in prompts, improve code quality or lead to unintended reproduction of technical debt. The findings reveal that while Copilot can sometimes generate solutions that address TODO comments, it also risks perpetuating technical debt symptoms. The authors propose best practices in prompt engineering to reduce these risks, showing that modifying prompt content can enhance code quality and mitigate technical debt issues.

While the studies by Zampetti et al. \cite{Zampetti-2020} and O'Brien et al. \cite{obrien2024prompt} are about SATD repayment, their objectives differ from ours. Zampetti et al.’s approach focuses on predicting the category of changes required to address SATD, whereas our aim is to generate code that directly addresses SATD. OBrien et al.'s approach relies on the method's signature and docstring (with or without an injected SATD comment) to generate the method body, ignoring the ground truth code. In other words, their approach disregards the current implementation of the method, and during the evaluation phase, they did not consider the ground truth code from the repository. Instead, they manually checked whether the generated code addressed the SATD, regardless of whether it was the expected code and had the expected functionality for that project. The only prior work that accounted for ground truth from real SATD repayment scenarios is the study by Mastropaolo et al. \cite{Mastropaolo-2023}, with which we compare our results in this paper.

\subsection{Other Studies on SATD Management}

SATD management encompasses tasks such as identification, classification, and tracking. Identification methods determine if a comment signals technical debt (SATD), with approaches ranging from rule-based (e.g., MAT~\cite{Guo-2021}, PENTACET~\cite{Sridharan-2023}) to machine learning (ML) models like CNN~\cite{Ren-2019}, BERT~\cite{Prenner-2022}, and LLMs \cite{Sheikhaei-2024}. Classification, introduced by Maldonado and Shihab~\cite{Maldonado-2015}, categorizes SATD comments (e.g., Design, Requirement), with studies expanding this categorization, especially in domains like Blockchain and ML~\cite{Pinna-2023,OBrien-2022,Bhatia-2023}. 

SATD tracking, which builds on SATD identification, focuses on tracing each SATD's lifespan across a repository's commit history, providing essential data on SATD removals for training automated SATD repayment models. Maldonado et al.\cite{maldonado2017empirical} introduced a basic approach that tracks SATDs across file versions by matching content, labeling SATDs as deleted if they no longer match. However, this method fails when multiple SATDs share content or when minor edits occur, causing tracking errors. AlOmar et al.\cite{AlOmar-2022} improved upon this with SATDBailiff, which uses method and class names to map SATDs more accurately across commits. SATDBailiff marks content changes as SATD\_CHANGED actions, though its Java parser dependency and limitations with multiple SATDs in the same method affect adaptability and accuracy. Addressing these limitations, Sheikhaei et al. proposed SATD Tracker~\cite{Sheikhaei-2023}, which enhances SATD tracking by storing and updating the current location (file and line number) of all detected SATDs through all commits. It uses the diff between every consecutive pair of commits to track detected SATDs until they are deleted or persist to the last commit. In this paper, we use the SATD Tracker because it can be applied to any programming language and is much faster than SATDBailiff, as it does not require a parser to track SATD locations. 

\subsection{Benchmarking LLMs on SE Tasks}

LLMs have been successfully employed in different Software Engineering (SE) tasks. Hou et al.~\citep{Hou-2024} conducted a systematic literature review on LLM4SE by studying 229 research papers from 2017 to 2024. They reported that the existing studies covered six stages of software engineering from requirement to software maintenance and management. Software development leads with approximately 56.6\%, and focuses on tasks such as code generation, code completion, code translation, and API documentation. Chen et al. notably posed \textit{HumanEval} to validate the functional correctness of generated code by creating a set of manually curated problems \cite{Chen2021EvaluatingLL}. Cassano et al. amplified \textit{HumanEval} to create \textit{MultiPL-E} for code translation \cite{Cassano-2023}. Software maintenance follows by 22.7\%, with tasks like program repair, code clone detection, and debugging that help maintain code quality over time. Software quality assurance represents 15.1\%, supporting tasks such as vulnerability detection, test generation, and defect detection to enhance testing processes. Requirements engineering constitutes roughly 3.9\%, addressing tasks such as requirement classification and traceability automation to improve requirements analysis. Software design covers around 0.9\%, with tasks like rapid prototyping and GUI retrieval, though this area is less explored. Finally, software management represents approximately 0.7\%, including tasks like effort estimation and tool configuration, showing potential for aiding in project planning and resource management.

\section{Empirical Study Design}\label{sec:research-questions}
In this section, we present our approach to addressing the proposed four research questions.

\subsection{RQ1- \rqone}\label{subsec:RQ1}

To create a SATD repayment dataset from real-world projects, we build upon the methodology of Mastropaolo et al.~\cite{Mastropaolo-2023}. Similar to them, we select GitHub repositories that meet the following criteria: at least 500 commits, 10 contributors, 10 stars, and no forks.To identify repositories that meet these criteria, we use SEART~\cite{Dabic-2021}, a tool capable of retrieving metadata for GitHub repositories based on commonly used selection criteria. Unlike Mastropaolo et al., who focused exclusively on Java repositories and relied on SATDBailiff for SATD extraction, we adopt SATD Tracker~\cite{Sheikhaei-2023}, a language-independent tool, to extract SATD items along with their lifespan from the main branches of both Python and Java repositories on GitHub.

Once SATD items were extracted from the target repositories, Mastropaolo et al.~\cite{Mastropaolo-2023} applied eight filtering steps to create a clean dataset of SATD repayments. We follow their approach and include these eight filtering steps in our pipeline. The steps are as follows:

\begin{enumerate}
    \item \textbf{The SATD should have been deleted, indicating a possibility of SATD repayment.} Based on the commit in which the SATD is deleted, we extract the code before and after the SATD removal.
    \item \textbf{The SATD should consist of three or more words.} Short SATDs such as ``// TODO" do not provide enough description for the models to automatically repay them.
    \item \textbf{The SATD should be defined inside a method or immediately above it.} Addressing class-level or file-level SATD is more challenging than method-level SATD because a broader range of code needs to be processed and edited to repay them. Since Mastropaolo et al. reported that automated SATD repayment is a challenging task even for method-level SATDs, we focus on method-level SATDs and leave class-level and file-level SATDs for future work.
    \item \textbf{The containing method's name still exists after repayment.} This constraint ensures that the SATD removal is not due to method or class removal. Zampetti et al. \cite{Zampetti-2018} discovered that 33\% to 63\% of method-level SATD removals resulted in a code change, meaning the containing method was updated. For the remaining instances, removal was due to class deletion, method deletion, or SATD comment removal while leaving the method code unchanged.
    \item \textbf{The containing method is updated after repayment.} This constraint filters out items where the SATD comment is removed without modifying the code.
    \item \textbf{Remove duplicates.} Based on the input (the method that includes the SATD) and the output (the method in which the SATD no longer exists), we remove duplicates.
    \item \textbf{Remove items containing non-ASCII letters.} This eliminates items with non-English comments.
    \item \textbf{Remove items where the number of tokens before or after SATD repayment exceeds 1,024.} As models are restricted by their input/output window size, this is a common practice in the software engineering literature~\cite{Mastropaolo-2023, macedo2024exploring}. 
\end{enumerate}

Although the filtering steps described above are necessary to create a dataset of SATD repayments, they are not sufficient to produce a clean dataset. The first issue is that there could be multiple SATDs in the input code, or there may exist some SATDs in the output code. Having multiple SATDs in the input code makes it difficult to distinguish which code updates are related to which SATD removal. Additionally, the presence of SATDs in the output code raises the suspicion that the input SATD is being converted into another SATD, with or without partial repayment. To address this issue, we add the following filter:

\begin{itemize}
    \item \textbf{(9) No other SATDs exist in the method before repayment. Also, no SATDs exist in the method after repayment.}
\end{itemize}

The second issue, as explained in the introduction, is that although the SATD is removed in the specified commit and the code is updated, we found that in 38 out of 100 randomly selected samples from the Java dataset and 55 out of 100 from the Python dataset, the code updates are not relevant to SATD repayment. Therefore, we add the following filter as the last step in the pipeline:

\begin{itemize}
    \item \textbf{(10) Remove items where the method update is not related to SATD repayment.}
\end{itemize}

To apply this filter, we leverage the Llama-3-70B model and use a zero-shot prompt to detect if the code update is relevant to SATD repayment. Figure~\ref{fig:prompt-for-last-filter} shows our prompt for this filter. In this prompt, we use the Chain-of-Thought (CoT) approach by asking the model to begin by identifying the code updates in the second version (where the SATD is removed) compared to the first version (which contains the SATD). Then, the model is asked to check if the code updates are mainly intended to address the issue mentioned in the SATD comment. We allow the model to answer ``Unclear'' if it cannot clearly identify the purpose of the code changes or the SATD comment. Once we receive the model's answer, we use a heuristic code to extract the label from the generated response. We treat the ``Unclear'' label as ``No'' to ensure dataset cleanliness.

\begin{figure}[h!]
    \centering
    \begin{tcolorbox}[colback=gray!5!white, colframe=gray!80!black, title=Prompt Template]
        Two versions of a method are provided below. The first version contains a Self-Admitted Technical Debt (SATD) comment, while the SATD comment no longer exists in the second version. Analyze if the code updates in the second version are related to resolving that SATD, considering the surrounding code context in addition to the SATD comment itself.
        
        \vspace{0.5em}
        \#\#\# Version 1:\\
        \{containing\_method\_before\_repayment\}

        \vspace{0.5em}
        \#\#\# Version 2:\\
        \{containing\_method\_after\_repayment\}

        \vspace{0.5em}
        \#\#\# SATD comment:\\
        \{SATD\_comment\}

        \vspace{0.5em}
        \#\#\# Consider the following questions in your analysis:\\
        - Shortly explain what specific changes were made in Version 2 compared to Version 1?\\
        - Are these changes mainly proposed to address the issue mentioned in the SATD comment? Please answer with Yes, No, or Unclear.
    \end{tcolorbox}
    \caption{Prompt used in final filtering step to detect relevance of code updates to SATD repayment}
    \Description{}
    \label{fig:prompt-for-last-filter}
\end{figure}

To evaluate the effectiveness of this approach, we use the aforementioned manually labeled 100 samples from each dataset and report the precision, recall, and F1 score of this approach on the ``Yes'' class for each dataset separately. The ``Yes'' class represents the items where the changes after SATD deletion are mainly proposed to address the issue mentioned in the SATD comment.


\subsection{RQ2- \rqtwo}\label{subsec:RQ2}

To evaluate the quality of generated code for SATD repayment, we consider three existing evaluation metrics: Exact Match (EM), BLEU, and CrystalBLEU. EM and BLEU are widely adopted in the software engineering domain for assessing generated text and code~\cite{Hou-2024}. We also include CrystalBLEU~\cite{Eghbali-2023}, as it, along with EM, was used by Mastropaolo et al.~\cite{Mastropaolo-2023} in their evaluation of automated SATD repayment models.

In the following, we introduce each of these evaluation metrics, identify their limitations, and propose solutions to address them. We then present our new metric, Line-Level Exact Match on Diff (LEMOD), which overcomes the shortcomings of the existing approaches. Unlike EM, LEMOD avoids underestimating model performance, and unlike BLEU and CrystalBLEU, it is intuitive, easy to interpret, and specifically tailored to measure the accuracy of code changes relevant to SATD repayment.

\vspace{0.5em}
\noindent \textbf{Exact Match (EM):} Given the generated code (a method) for addressing a SATD, if the whole code exactly matches the ground truth, i.e., the method updated by the developer through a commit, return 1; otherwise, return 0. This approach is easy to understand, implement, and interpret. The main issue with this method is that it does not score partial similarities. Even worse, if the difference is just whitespace or a code comment, it treats that as an unmatched pair. To mitigate this issue, we apply three preprocessing steps to the generated code and the ground truth before checking for an exact match.

\begin{enumerate}
    \item \textbf{Remove import statements:} All import statements should be removed. Since the ground truth is an extracted method from a Python or Java file and does not contain the lines that import the necessary packages, we need to remove the import statements from the generated code.
    \item \textbf{Remove code comments:} Code comments, including inline comments, Python Docstrings, and Javadocs, are removed from generated code and ground truth, as they do not affect program functionality. In addition, when asked to modify the code, LLMs often add comments or explanations that describe the changes made, which can bias the performance evaluation. 
    \item \textbf{Reformat the code:} Finally, we reformat the code to ensure that whitespace differences do not negatively affect performance. For Python code, we use the Black package to automatically reformat the code to follow the PEP 8 style guide and ensure consistency across different sources. For Java code, since indentation does not affect syntax, we simply remove all whitespace.
\end{enumerate}

Once we have applied the above preprocessing steps, the code is ready for exact match checking.

\vspace{0.5em}
\noindent \textbf{BLEU:} The BLEU (Bilingual Evaluation Understudy) score is a metric used to evaluate the quality of machine-translated text by comparing it to one or more reference translations. It calculates the overlap of n-grams (sequences of words) between the machine-generated translation and the reference texts, with higher scores indicating greater similarity. The BLEU score formula is:

\begin{equation}
\text{BLEU} = \text{BP} \times \exp \left( \sum_{n=1}^{N} w_n \log p_n \right)    
\end{equation}

\noindent where \( N \) represents the highest n-gram order considered (typically 4, making this BLEU-4 that uses unigrams, bigrams, trigrams, and four-grams). In this study, we use BLEU-4, with each n-gram order contributing equally (\( w_n = 0.25 \)). The term \( P_n \) is the precision for the \( n \)-grams, measuring the proportion of \( n \)-grams in the candidate translation that match those in the reference. A brevity penalty (BP) is applied to penalize overly short translations and is defined as:

\begin{equation}
\text{BP} = 
\begin{cases} 
      1 & \text{if } c > r \\
      \exp \left(1 - \frac{r}{c}\right) & \text{if } c \leq r 
\end{cases}
\end{equation}

\noindent where \( c \) is the length of the candidate translation, and \( r \) is the length of the reference translation. This penalty ensures that translations are not only accurate in terms of content but also similar in length to the reference. The BLEU score ranges from 0 to 1, with 1 indicating a perfect match.

Before calculating the BLEU score, we remove imports, comments, and docstrings/Javadoc (ICD) to ensure they do not adversely affect the score.

\vspace{0.5em}
\noindent \textbf{CrystalBLEU:} CrystalBLEU~\cite{Eghbali-2023} is a modification of the BLEU metric designed to reduce noise caused by trivially shared n-grams, which often result from the syntactic verbosity and coding conventions of programming languages. CrystalBLEU addresses this issue by excluding the most common n-grams across the dataset when calculating the BLEU score. As with BLEU, we remove ICD before calculating CrystalBLEU.

\begin{figure}[h]
  \centering
  \includegraphics[width=\linewidth]{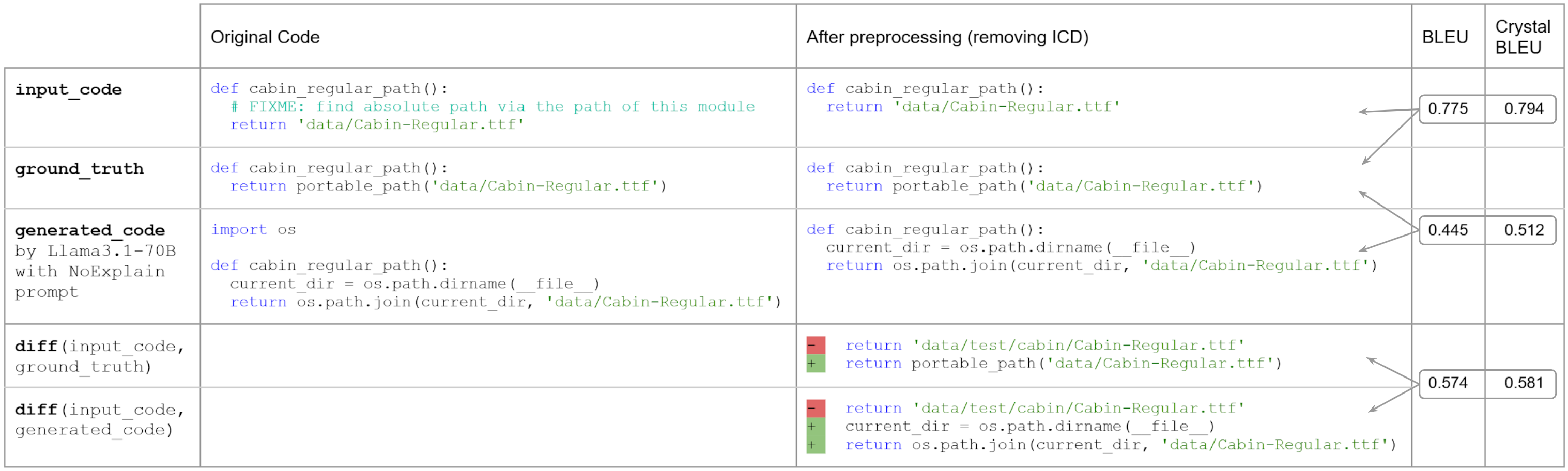}
  \caption{Comparison of BLEU and CrystalBLEU scores on the whole code versus on the diffs.}
  \label{fig:BLEU-diff}
\end{figure}

\vspace{0.5em}
\noindent \textbf{BLEU-diff} and \textbf{CrystalBLEU-diff}: While BLEU and CrystalBLEU are effective for evaluating text and code similarity, directly applying these metrics to the entire generated code for SATD repayment may yield misleading results. For example, as shown in Figure~\ref{fig:BLEU-diff}, simply removing SATD comments can result in higher similarity scores following the whole-code based evaluation. In contrast, a solution that aims to address the SATD receives lower similarity scores, though it may differ from how developers implemented it.

To overcome this limitation, we propose a diff-based evaluation approach. Rather than evaluating the whole code, we compute the differences (diffs) between the code before repayment and the ground truth, and the code before repayment and the generated code. We then calculate BLEU or CrystalBLEU scores on these diffs. This method enables us to focus exclusively on code updates, ignoring unchanged lines that do not contribute to SATD repayment. We refer to these modified metrics as \textit{BLEU-diff} and \textit{CrystalBLEU-diff}. To identify inserted and deleted lines, we use the \textit{difflib} package~\cite{difflib}.

Figure~\ref{fig:BLEU-diff} illustrates the two evaluation approaches with an example. The first three rows present the input code (containing the SATD), the ground truth code (after SATD repayment), and the generated code produced by the Llama-3.1-70B model using the NoExplain prompt template (ref. Table~\ref{tab:prompt-templates-for-SATD-repayment}). We also present the original source code and one after preprocessing. The last two rows display the two code diffs, the \textit{reference diff}, i.e., the changes between the input code and the ground truth code, and the \textit{candidate diff}, i.e., the changes between the input code and the generated code. The BLEU and CrystalBLEU scores are shown in the corresponding columns. The first two scores for each metric are computed by applying BLEU and CrystalBLEU to the whole code. Notably, these scores are higher for the input code than for the generated code, despite the input code not addressing the SATD. In contrast, when BLEU and CrystalBLEU are applied to the diffs (last two rows), the BLEU-diff and CrystalBLEU-diff scores for the generated code are 0.574 and 0.581, respectively. Since the BLEU-diff and CrystalBLEU-diff scores for the input code are always zero (due to the resulting empty string when applying the diff), the generated code consistently achieves a higher score than the input code. While this table highlights the limitations of BLEU and CrystalBLEU using an extreme example, we will explore this issue in greater detail in RQ2.

\noindent \textbf{LEMOD:} While BLEU-diff and CrystalBLEU-diff address the limitations of whole-code matching by focusing on code changes, their reported values remain difficult to interpret. They rely on complex calculations involving n-gram overlaps and brevity penalties, making them less intuitive than simpler metrics like EM. To address this, we propose Line-Level Exact Match on Diff (LEMOD)-a new evaluation approach that is easier to implement, more interpretable, and still offers finer-grained insights into SATD repayment quality. In LEMOD, we focus on the line-level changes between the input code, ground truth, and generated code. The approach computes precision, recall, and F1 score based on the overlap of modified lines between two code diffs, i.e., \textit{reference\_diff} and \textit{candidate\_diff}.

\begin{itemize}
    \item \textbf{reference\_diff:} set of lines in diff(input\_code, ground\_truth)
    \item \textbf{candidate\_diff:} set of lines in diff(input\_code, generated\_code)
    \item \textbf{LineP} = count(intersection(reference\_diff, candidate\_diff)) / count(candidate\_diff)
    \item \textbf{LineR} = count(intersection(reference\_diff, candidate\_diff)) / count(reference\_diff)
    \item \textbf{LineF} = $2 \times LineP \times LineR / (LineP + LineR)$
\end{itemize}

For example, consider the input code (before addressing SATD), ground truth code, and generated code shown in Figure~\ref{fig:BLEU-diff}. The computed LEMOD metrics are as follows: LineP = 1/3, LineR = 1/2, and LineF = 0.4, indicating that 33\% of the model's proposed changes were relevant (precision), 50\% of the ground truth changes were captured (recall), and the overall quality of the updates balances at 0.4 (F1 score).

To check the effectiveness and explainability of these evaluation metrics, we employ four open LLMs (Llama-3.1-8B-Instruct, Llama-3.1-70B-Instruct, Gemma-2-9B-it, and DeepSeek-Coder-V2-Lite-Instruct) and one closed model, GPT4o-mini. The rationale for selecting these models is provided in Section~\ref{subsec:selected-LLMs}.

We consider four different prompts for each model:

\begin{itemize}
    \item \textbf{Mastropaolo-T2:} The second template used by Mastropaolo et al. when employing GPT-3.5, which achieved higher performance compared to their T1 and T3 templates.
    \item \textbf{NoExplain:} A template that uses ``\#\#\#'' separators for each section of the prompt, instructing the model to provide only the output code without any explanation.
    \item \textbf{CoT1:} A chain of thought (CoT) prompt that asks the model to first explain how to resolve the SATD and then generate the output code. Similar to NoExplain, ``\#\#\#'' separators are used for each section of the prompt.
    \item \textbf{CoT2:} A slight modification of CoT1, designed to test if the performance metrics are robust to minor changes in the prompt.
\end{itemize}

These four prompt templates are shown in Table~\ref{tab:prompt-templates-for-SATD-repayment}.

\begin{table}
\caption{Zero-shot prompt templates for SATD repayment task}
\label{tab:prompt-templates-for-SATD-repayment} 
\resizebox{15cm}{!}{
\begin{tabular}{|p{0.15\linewidth}|p{0.85\linewidth}|}
\hline
\textbf{Template name} & \textbf{Template structure} \\
\hline
Mastropaolo-T2 & \texttt{Perform removal of this SATD: \{comment\} from this code: \{code\}} \\
\hline
NoExplain & \texttt{How to update the following code to resolve the SATD? No need to explain. Just provide the updated code.} \\
& \texttt{\#\#\# Code:} \\
& \texttt{\{code\}} \\
& \texttt{\#\#\# SATD comment:} \\
& \texttt{\{comment\}} \\
& \texttt{\#\#\# Updated code after SATD repayment:} \\
\hline
CoT1 & \texttt{How to update the following code to resolve the SATD?} \\
& \texttt{\#\#\# Code:} \\
& \texttt{\{code\}} \\
& \texttt{\#\#\# SATD comment:} \\
& \texttt{\{comment\}} \\
& \texttt{\#\#\# Consider the following questions in your answer:} \\
& \texttt{Shortly explain how to resolve the SATD.} \\
& \texttt{Provide the updated code.} \\
\hline
CoT2 & \texttt{How to update the following code to resolve the Self-Admitted Technical Debt (SATD)?} \\
& \texttt{\#\#\# Code:} \\
& \texttt{\{code\}} \\
& \texttt{\#\#\# SATD comment:} \\
& \texttt{\{comment\}} \\
& \texttt{\#\#\# Consider the following questions in your answer:} \\
& \texttt{1. Briefly explain how to resolve the SATD.} \\
& \texttt{2. Provide the updated code.} \\
\hline
\end{tabular}
}
\end{table}

With five models and four prompt templates, we need to test 20 model-template pairs across our two datasets, resulting in 40 experiments for RQ2. To expedite the tests, we randomly select 1,000 samples from each dataset as our test set for RQ2.

Since the validity of the exact match (EM) approach is well-established, any effective evaluation measure should have a high correlation with it. Therefore, we report the correlation between all other metrics and the exact match to assess their alignment with EM.


\subsection{RQ3- \rqthree}\label{subsec:RQ3}

In RQ1, we proposed two novel filtering steps to create a cleaner SATD repayment dataset. In RQ3, we aim to apply these filters to the Mastropaolo dataset to investigate how a cleaner dataset impacts the performance of LLM-based automated SATD repayment approaches, using the evaluation metrics defined in RQ2. By reducing the likelihood of including non-SATD repayments, we hypothesize that the filtered dataset will lead to enhanced performance in automated SATD repayment tasks.

Mastropaolo et al. evaluated two types of approaches: (1) GPT-3.5-Turbo, with three prompt templates (T1, T2, T3), and (2) four fine-tuned models (M0 to M3) based on the CodeT5-base model, each trained on different datasets. In RQ3, we extend their approach by considering both a GPT model without fine-tuning and a fine-tuned model. We select GPT-4o-mini due to its significantly lower cost compared to GPT-3.5-Turbo and its superior performance across multiple benchmarks\footnote{See https://openai.com/index/gpt-4o-mini-advancing-cost-efficient-intelligence/}. Experiments are conducted using GPT-4o-mini on both the original Mastropaolo testing dataset and its cleaned version. Furthermore, based on our RQ2 findings with the Java dataset, i.e., our NoExplain template outperformed their best-performing template T2, we adopt the NoExplain template for this experiment.

For the fine-tuned model, we select the M1 approach for its simplicity but replace CodeT5-base with CodeT5p-770m, a larger and more advanced model. To assess the impact of using a cleaner training and testing dataset, we fine-tune CodeT5p-770m on the filtered training set of the Mastropaolo dataset and evaluate it on the corresponding filtered test set. Since filtering may influence the performance of fine-tuned models, we ensure a fair comparison by creating controlled training and evaluation sets. These are generated by randomly sampling from the original training and evaluation data to match the size of the filtered sets.


\subsection{RQ4- \rqfour}\label{subsec:RQ4}

Mastropaolo et al. reported that fine-tuning smaller models outperformed prompt-based LLMs. However, this conclusion may not hold due to the presence of non-SATD repayments in their dataset and the limitations of the evaluation metrics they used. Additionally, it remains unclear whether this performance trend generalizes to other programming languages, such as Python. Therefore, in RQ4, we revisit this comparison using our newly created benchmarks (Java and Python datasets) and the improved evaluation metrics introduced in RQ2.

To implement the prompt-based approach, we evaluate the same five LLMs previously considered in RQ2. For each model, we select the best-performing prompt template based on its performance across four evaluation metrics: EM, BLEU-diff, CrystalBLEU-diff, and LineF1 (measured using LEMOD). When there is a conflict, i.e., where BLEU-diff and CrystalBLEU-diff favor one template, while EM and LineF1 favor another, we resolve this by prioritizing the template preferred by EM and LineF1, given their complementary nature and interpretability.

For the fine-tuning-based approach, we use the CodeT5p family~\cite{Wang-2023-codet5p}, an enhanced version of the original CodeT5 model used in Mastropaolo's study. Released in October 2023, CodeT5p (CodeT5+) incorporates additional pretraining tasks, such as span denoising, causal language modeling, contrastive learning, and text-code matching, allowing it to learn richer representations. While Mastropaolo et al. used only the base variant (220 million parameters), we expand the investigation to include two CodeT5p variants: the 220-million and the 770-million parameter models, enabling us to analyze the impact of model size on SATD repayment performance.

Since fine-tuning requires training data, we split each dataset (Java and Python) into training, validation, and test sets using a repository-wise split to avoid data leakage. Specifically, we randomly allocate 85\% of repositories to the training set, 5\% to the validation set, and 10\% to the test set. This approach ensures that no code from the same repository appears in multiple subsets, preserving the integrity of the evaluation. Table~\ref{tab:train-val-test-count} summarizes the number of SATD samples allocated to each subset for the two datasets. To evaluate the prompt-based approaches, we use only the test portion of each dataset.

\begin{table} [h]
  \caption{Number of SATD samples in training, validation, and test sets for each dataset}
  \label{tab:train-val-test-count}
  \begin{tabular}{lllll}
    \toprule
    & Train & Validation & Test & Overall \\
    \midrule
    Python dataset & 50,075 & 2,902 & 5,745 & 58,722 \\
    Java dataset & 83,211 & 5,123 & 9,013 & 97,347 \\
  \bottomrule
\end{tabular}
\end{table}

\section{Study Setup}\label{sec:study-setup}
In this section, we summarize the study setup for the experiments involved in addressing four RQs.

\subsection{Datasets}\label{subsec:datasets}

In this study, we follow the approach introduced in RQ1 to create two SATD repayment datasets: one for Python and one for Java. These two datasets are employed in RQ2 and RQ4. For RQ3, we use the Mastropaolo dataset. Table~\ref{tab:datasets-statistics} presents basic statistics for the three datasets. Although the number of processed repositories in our Java dataset is approximately ten times greater than in the Mastropaolo dataset, the number of identified \textit{SATD repayment samples} is nearly twenty times higher than in the Mastropaolo dataset, primarily due to their failure to process large repositories.

\begin{table} [h]
  \caption{Basic statistics of the Maldonado dataset and our Java and Python datasets}
  \label{tab:datasets-statistics}
  \begin{tabular}{lccccc}
    \toprule
    & Total & Number & SATD & Processed & Repositories having \\
    & number & of & repayment & repositories & at least one SATD \\
    & of SATDs & filters & samples &  & in the filtered dataset \\
    \midrule
    Our Python dataset & 1,607,408 & 10 & 58,722 & 13,748 & 7,219 \\
    Our Java dataset   & 2,672,485 & 10 & 97,347 & 7,694  & 5,181 \\
    Mastropaolo dataset (Java) & Unknown & 8 & 5,039 & 809 & 596 \\
  \bottomrule
\end{tabular}
\end{table}

\subsection{Evaluation Metrics}\label{subsec:evaluation-metrics}

In RQ2, we introduce three metrics for evaluating SATD repayment: EM, BLEU, and CrystalBLEU. Additionally, we present modified versions of these metrics, i.e., BLEU-diff and CrystalBLEU-diff, to address issues identified with BLEU and CrystalBLEU. Moreover, we introduce a new evaluation metric, Line-Level Exact Match on Diff (LEMOD). LEMOD provides a finer-grained evaluation by reporting precision, recall, and F1 score based on line-level comparisons of code diffs, offering greater interpretability and ease of implementation compared to BLEU and CrystalBLEU.
We report results across all six metrics to enable a comprehensive comparison and analysis of evaluation metrics. For RQ3 and RQ4, we focus on four key metrics: EM, BLEU-diff, CrystalBLEU-diff, and LEMOD, as these metrics provide a balanced evaluation of correctness, code similarity, and the quality of generated updates.

\subsection{Selected Large Language Models}\label{subsec:selected-LLMs}

To select the large language models for our research questions, we referred to the Chatbot Arena leaderboard \cite{lmarena_ai}. At the time we were cleaning the dataset (RQ1), the best-performing open LLM was \textbf{Llama-3-70B-Instruct}. Therefore, we used that model to apply the last step of our filtering pipeline.

For prompt-based SATD repayment (RQ2 to RQ4), we selected these five models according to their high scores in the Chatbot Arena leaderboard (as of July 31, 2024):

\begin{itemize}
    \item \textbf{GPT-4o-mini-2024-07-18:} This model, along with GPT-4o-2024-05-13 and Claude 3.5 Sonnet, ranked first in the Chatbot Arena leaderboard. As GPT-4o-mini provides the lowest API price, we selected it as the proprietary model in our experiments.
    \item \textbf{Llama-3.1-70B-Instruct:} Ranked first among open models in the Chatbot Arena (Code section).
    \item \textbf{Gemma-2-9B-it:} Ranked first among open models with less than 10 billion parameters in the Chatbot Arena (Code section).
    \item \textbf{Llama-3.1-8B-Instruct:} Ranked second among open models with less than 10 billion parameters in the Chatbot Arena (Code section).
    \item \textbf{DeepSeek-Coder-V2-Lite-Instruct:} A model specifically created for coding. We use the Lite version to be able to run it on our infrastructure.
\end{itemize}

For the fine-tuning experiments in RQ3 and RQ4, we use models from the CodeT5p family (also known as CodeT5+), which is an enhanced version of the original CodeT5 models employed by Mastropaolo et al.~\cite{Mastropaolo-2023}.

\subsection{Implementation}\label{subsec:implementation}

For all LLMs under study, we downloaded their official checkpoints from Hugging Face and used the default 16-bit or 32-bit precision mode for inference. An exception was made for the Llama-3-70B-Instruct model (used in RQ1), where we employed its quantized AWQ version to enable it to run on a single Nvidia RTX A6000 GPU with 48 GB of VRAM. All other experiments were conducted on a single A6000 GPU, except for the Llama-3.1-70B-Instruct model, which required four A6000 GPUs for execution.

To fine-tune the CodeT5p models, we utilized the replication package provided by Mastropaolo et al.~\cite{Mastropaolo-2023}. To optimize performance, we experimented with different learning rates while keeping all other hyperparameters unchanged. Based on these experiments, we selected the best-performing learning rates: 1e-5 for CodeT5p-220m and 5e-6 for CodeT5p-770m, which were used for reporting the final results.

For prompt-based experiments, we utilized vLLM~\cite{kwon2023efficient}, an open-source, high-performance inference engine for serving large language models. vLLM optimizes memory usage and inference speed by batching prompts, ensuring efficient execution. To ensure deterministic outputs and enhance reproducibility, we set the temperature to 0, guaranteeing that identical inputs produce consistent outputs. The maximum output length (\verb|max_tokens|) was set to 2,048 tokens, allowing the models sufficient room to generate complete answers.

\section{Results and Analysis}\label{sec:results}
\subsection{RQ1 \rqone}\label{subsec:results-rq1}

Table~\ref{tab:dataset-creation-process} summarizes the statistics for each step in our data creation process. Using the SEART tool, we identified 14,097 Python repositories and 7,982 Java repositories that met our selection criteria. The SATD Tracker tool successfully processed 97.5\% of the Python projects and 96.4\% of the Java projects, with the remaining repositories excluded due to failures in a third-party Python package for parsing Git diff outputs. We detected 1,059,299 SATD items in Python repositories and 2,672,485 in Java repositories. Rows 4–13 in the table outline our 10-step filtering approach to create a clean dataset of SATD repayments, with two newly proposed filtering steps highlighted in rows 9 and 13.

\begin{table}[h]
  \caption{Dataset creation process for Python and Java: SATD extraction (rows 1–3) and filtering pipeline (rows 4–13)}
  \label{tab:dataset-creation-process}
  \resizebox{15cm}{!}{
  \begin{tabular}{p{0.03\linewidth}p{0.74\linewidth}p{0.12\linewidth}p{0.11\linewidth}}
    \toprule
    Row & Our dataset creation steps & Python & Java \\
    \midrule
    1 & Repositories having at least 500 commits, 10 contributors, 10 stars, and not being forked & 14,097 & 7,982 \\
    2 & Number of repositories that our SATD-Tracker successfully extracted SATDs & 13,748 (97.5\%) & 7,694 (96.4\%) \\
    3 & Total number of SATDs & 1,607,408 & 2,672,485 \\
    4 & Number of SATDs that are deleted (potentially repaid) & 1,059,299 & 1,847,392 \\
    5 & Number of SATDs after removing those with length of two or less words & 949,188 & 1,623,408 \\
    6 & Number of SATDs that are inside methods & 723,258 & 1,216,775 \\
    7 & The containing method’s name still exist after repayment & 325,031 & 507,287 \\
    8 & The containing method is updated after repayment & 288,007 & 423,915 \\
    9 & No other SATDs existed in that method before repayment. Also, no SATD exists in the method after repayment. & 171,825 & 275,883 \\
    10 & Number of SATDs after removing duplicates & 143,341 & 199,219 \\
    11 & Number of SATDs after removing those having non-ASCII & 140,929 & 196,741 \\
    12 & The number of tokens in before and after SATD repayment is less than 1,024 & 131,945 & 187,380 \\
    13 & Use LLM to remove items that the method update is not related to SATD repayment & 58,722 & 97,347 \\
  \bottomrule
\end{tabular}
}
\end{table}

We ended up with 58,722 and 97,347 SATD repayment samples for Python and Java, respectively. For each sample, the dataset provides various information, including the GitHub user, the project (repository name), the file path that contains the SATD, the creation commit and deletion commit of the SATD, the line number of the SATD when it was created, and also its line number just before the deletion commit. Using this information, for all extracted SATD items that have a deletion commit, we processed the containing file in the deletion commit and its parent to extract the \verb|containing_method_before_repayment| and \verb|containing_method_after_repayment| fields. The first one is used as the input for the LLMs, and the second one is used as the ground truth in our experiments for RQ2 to RQ4. In RQ1, to apply the last filtering step, we provided both fields and asked the model to determine if the changes made in the second code address the issue mentioned in the SATD comment.

Table~\ref{tab:rq1-results} presents the precision, recall, and F1 score of our LLM-based approach for identifying code updates related to SATD repayment, evaluated using 200 manually labeled candidate SATD repayments prior to applying the filtering step (see Section~\ref{sec:study-setup}). While we used the Llama-3-70B-Instruct model to filter the entire dataset in RQ1, we also tested GPT-4 as a baseline on the same manually labeled dataset, using the same prompt as for the Llama-3 model. As shown in Table~\ref{tab:rq1-results}, GPT-4 achieves a higher F1 score across both datasets. However, the Llama-3 model exhibits higher precision, a critical factor for generating a clean dataset. Thus, our final dataset for RQ2–RQ4 was collected using the Llama-3-based filtering approach.

\begin{table}
  \caption{Evaluation of our approach in identifying items whose code updates are related to SATD repayment (``Yes'' category)}
  \label{tab:rq1-results}
  \begin{tabular}{llllll}
    \toprule
    Dataset & \#items in Yes category & Model & Precision & Recall & F1 \\
    \midrule
    Python & 45 out of 100 & TechxGenus/Meta-Llama-3-70B-Instruct-AWQ & 0.830 & 0.867 & 0.848 \\
    & & GPT4 (baseline) & 0.800 & 0.978 & 0.880 \\
    \hline
    Java & 62 out of 100 & TechxGenus/Meta-Llama-3-70B-Instruct-AWQ & 0.926 & 0.806 & 0.862 \\
    & & GPT4 (baseline) & 0.881 & 0.952 & 0.915 \\
  \bottomrule
\end{tabular}
\end{table}

\begin{tcolorbox}[enhanced,width=6in,size=fbox,drop shadow southwest,sharp corners]

\textit{RQ1 Summary: Using our SATD Tracker tool, we successfully extracted SATDs from 97.5\% of the targeted Python repositories and 96.4\% of the targeted Java repositories. These success rates are significantly higher than the 12\% success rate reported in the previous study by Mastropaolo et al. Our LLM-based approach, utilizing the quantized Llama-3-70B model, achieved an F1 score of 0.848 for Python and 0.862 for Java in identifying true SATD repayment samples, resulting in a less noisy dataset.} 

\end{tcolorbox}


\subsection{RQ2 \rqtwo}\label{subsec:results-rq2}

Table~\ref{tab:BLEU-CrystalBLEU-results} presents the BLEU and CrystalBLEU scores achieved by five LLMs on two sampled test sets, each containing 1,000 SATD repayments per programming language, evaluated using four distinct prompt templates and the whole-code based evaluation method. When comparing the scores of LLM-generated responses to the baseline (where the input code, after removing the ICD, is used as the generated code, i.e., ignoring the SATD comment rather than addressing it), we observe that most LLM-generated scores (48 out of 80 experiments) fall below the baseline. This observation aligns with the demonstration example in Figure~\ref{fig:BLEU-diff} and the hypothesis outlined in Section~\ref{subsec:RQ2}. In the few instances where LLM-generated scores surpass the baseline, the difference is minimal—less than 0.03, indicating only a marginal improvement. These findings confirm that measuring BLEU and CrystalBLEU based on the entire code is not suitable for evaluating automated SATD repayment.

\begin{table}
  \caption{BLEU and CrystalBLEU scores for five models each with four prompt templates on two datasets, compared to baseline input code (red underlined values < baseline)}
  \label{tab:BLEU-CrystalBLEU-results}
  \begin{tabular}{llllll}
    \toprule
    & & \multicolumn{2}{c}{Python} & \multicolumn{2}{c}{Java} \\
    \cline{3-4} \cline{5-6}
    Model & Prompt   & BLEU      & CrystalBLEU & BLEU      & CrystalBLEU \\
          & Template & WholeCode & WholeCode   & WholeCode & WholeCode \\
    \midrule
 Llama-3.1-8B-Instruct & Mastropaolo-T2 & 0.657 & 0.702 & 0.572 & \hl{0.610} \\
 & NoExplain & \hl{0.650} & 0.704 & 0.580 & 0.635 \\
 & CoT1 & \hl{0.627} & \hl{0.688} & \hl{0.521} & \hl{0.578} \\
 & CoT2 & \hl{0.625} & \hl{0.687} & \hl{0.515} & \hl{0.572} \\
 \hline
 Gemma-2-9B-it & Mastropaolo-T2 & 0.669 & 0.714 & 0.585 & 0.632 \\
 & NoExplain & \hl{0.624} & \hl{0.675} & 0.583 & 0.636 \\
 & CoT1 & \hl{0.511} & \hl{0.573} & \hl{0.469} & \hl{0.528} \\
 & CoT2 & \hl{0.513} & \hl{0.575} & \hl{0.483} & \hl{0.541} \\
 \hline
 DeepSeek-Coder-V2-Lite-Instruct & Mastropaolo-T2 & 0.658 & 0.699 & 0.572 & \hl{0.604} \\
 & NoExplain & 0.657 & 0.714 & 0.582 & 0.627 \\
 & CoT1 & \hl{0.634} & \hl{0.693} & \hl{0.559} & \hl{0.601} \\
 & CoT2 & \hl{0.624} & \hl{0.685} & \hl{0.539} & \hl{0.584} \\
 \hline
 Llama-3.1-70B-Instruct & Mastropaolo-T2 & \hl{0.643} & 0.698 & \hl{0.543} & \hl{0.583} \\
 & NoExplain & 0.666 & 0.723 & 0.572 & 0.621 \\
 & CoT1 & \hl{0.590} & \hl{0.649} & \hl{0.496} & \hl{0.543} \\
 & CoT2 & \hl{0.594} & \hl{0.653} & \hl{0.496} & \hl{0.545} \\
 \hline
 GPT-4o-mini-2024-07-18 & Mastropaolo-T2 & 0.660 & 0.706 & \hl{0.564} & \hl{0.597} \\
 & NoExplain & 0.673 & 0.730 & 0.595 & 0.642 \\
 & CoT1 & \hl{0.653} & 0.713 & \hl{0.550} & \hl{0.590} \\
 & CoT2 & \hl{0.642} & 0.701 & \hl{0.540} & \hl{0.581} \\
 \hline
 \multicolumn{2}{c}{Use the input code (after removing ICD) as a baseline} & 0.656 & 0.695 & 0.564 & 0.611 \\
  \bottomrule
\end{tabular}
\end{table}

Tables \ref{tab:rq2-python-results} and \ref{tab:rq2-java-results} summarize the performance of 20 experiments (five models with four prompt templates) conducted on the Python and Java datasets, respectively. Each model-prompt combination is evaluated using four metrics: EM, BLEU-diff, CrystalBLEU-diff, and LEMOD. LEMOD further provides line-level precision (LineP), recall (LineR), and F1 (LineF1) scores. The last three metrics are newly proposed in this study. 

As described in Section~\ref{subsec:RQ2}, since EM is intuitively a reliable metric for evaluating SATD repayment, we assess the effectiveness of all other metrics, including BLEU and CrystalBLEU, by calculating their correlation with EM. Table~\ref{tab:correlation-with-EM} presents the results of this correlation analysis. While BLEU and CrystalBLEU show minimal correlation with EM, all three of our proposed metrics exhibit a high correlation with it, indicating strong alignment with this intuitive metric. Among these, LineF1 achieves the highest correlation with EM.

\begin{table}
  \caption{EM, BLEU-diff, CrystalBLEU-diff, and LEMOD scores for five models each with four prompts on the \textbf{Python} dataset. For each metric, the maximum score across four prompts is bolded, and the maximum value across all 20 experiments is underlined.}
  \label{tab:rq2-python-results}
  \begin{tabular}{ll|ll|llllll}
    \toprule
    Model & Prompt   & Avg del. & Avg ins. & Exact & BLEU & CrystalBLEU & \multicolumn{3}{c}{LEMOD} \\
    \cline{8-10}
          & Template & lines    & lines    & Match & diff & diff        & LineP & LineR & LineF1 \\
    \midrule
 Llama-3.1-8B & Mastropaolo-T2 & 1.9 & 3.0 & 5.1 & 0.146 & 0.176 & 0.203 & 0.158 & 0.158 \\
 -Instruct & NoExplain & 3.2 & 4.5 & 7.7 & 0.298 & 0.360 & 0.387 & 0.306 & 0.306 \\
 & CoT1* & 6.2 & 6.0 & \textbf{9.1} & \textbf{0.336} & 0.411 & 0.383 & 0.360 & \textbf{0.329} \\
 & CoT2 & 7.3 & 5.6 & 7.1 & 0.333 & \textbf{0.419} & 0.365 & 0.367 & 0.324 \\ [0.5em]
 Gemma-2-9B & Mastropaolo-T2 & 1.6 & 3.1 & 7.3 & 0.198 & 0.226 & 0.293 & 0.205 & 0.216 \\
 -it & NoExplain* & 2.5 & 7.9 & \textbf{9.9} & 0.314 & 0.369 & 0.411 & 0.324 & \textbf{0.321} \\
 & CoT1 & 6.6 & 12.0 & 6.4 & \textbf{0.316} & \textbf{0.391} & 0.341 & 0.397 & 0.315 \\
 & CoT2 & 6.4 & 12.2 & 6.8 & \textbf{0.316} & 0.390 & 0.345 & 0.392 & 0.315 \\ [0.5em]
 DeepSeek- & Mastropaolo-T2 & 1.0 & 2.4 & 3.8 & 0.094 & 0.109 & 0.140 & 0.103 & 0.104 \\
 Coder-V2- & NoExplain & 3.9 & 3.4 & 5.6 & 0.256 & 0.321 & 0.315 & 0.257 & 0.252 \\
 Lite-Instruct & CoT1* & 7.7 & 4.5 & \textbf{7.7} & \textbf{0.296} & \textbf{0.368} & 0.328 & 0.331 & \textbf{0.290} \\
 & CoT2 & 8.5 & 4.7 & 5.8 & 0.287 & 0.364 & 0.313 & 0.339 & 0.281 \\ [0.5em]
 Llama-3.1-70B & Mastropaolo-T2 & 5.2 & 5.4 & \textbf{\underline{10.5}} & 0.328 & 0.393 & 0.373 & 0.350 & 0.325 \\
 -Instruct & NoExplain* & 5.9 & 4.7 & 10.0 & \textbf{\underline{0.378}} & \textbf{\underline{0.454}} & 0.419 & 0.402 & \textbf{\underline{0.373}} \\
 & CoT1 & 6.4 & 9.2 & 10.0 & 0.362 & 0.440 & 0.382 & 0.412 & 0.351 \\
 & CoT2 & 7.0 & 8.6 & 9.7 & 0.361 & 0.436 & 0.380 & 0.414 & 0.351 \\ [0.5em]
 GPT-4o-mini & Mastropaolo-T2 & 2.5 & 2.7 & 5.1 & 0.152 & 0.187 & 0.183 & 0.159 & 0.153 \\
 -2024-07-18 & NoExplain & 4.3 & 4.6 & 9.2 & 0.341 & 0.415 & 0.394 & 0.342 & 0.333 \\
 & CoT1* & 7.8 & 5.2 & \textbf{9.6} & \textbf{0.358} & \textbf{0.437} & 0.376 & 0.386 & \textbf{0.342} \\
 & CoT2 & 8.6 & 5.5 & 9.3 & 0.347 & 0.427 & 0.358 & 0.379 & 0.330 \\
  \bottomrule
\end{tabular}
\end{table}

\begin{table}
  \caption{EM, BLEU-diff, CrystalBLEU-diff, and LEMOD scores for five models each with four prompts on the \textbf{Java} dataset. For each metric, the maximum score across four prompts is bolded, and the maximum value across all 20 experiments is underlined.}
  \label{tab:rq2-java-results}
  \begin{tabular}{ll|ll|llllll}
    \toprule
    Model & Prompt   & Avg del. & Avg ins. & Exact & BLEU & CrystalBLEU & \multicolumn{3}{c}{LEMOD} \\
    \cline{8-10}
          & Template & lines    & lines    & Match & diff & diff        & LineP & LineR & LineF1 \\
    \midrule
 Llama-3.1-8B & Mastropaolo-T2 & 2.2 & 2.5 & 3.6 & 0.180 & 0.216 & 0.333 & 0.236 & 0.233 \\
 -Instruct & NoExplain* & 4.4 & 3.5 & \textbf{5.7} & 0.291 & 0.355 & 0.408 & 0.348 & \textbf{0.336} \\
 & CoT1 & 7.3 & 5.5 & 5.4 & \textbf{0.297} & \textbf{0.374} & 0.365 & 0.400 & 0.335 \\
 & CoT2 & 8.7 & 5.1 & 5.3 & 0.292 & 0.372 & 0.342 & 0.404 & 0.323 \\ [0.5em]
 Gemma-2-9B & Mastropaolo-T2 & 1.8 & 2.6 & 5.8 & 0.173 & 0.199 & 0.333 & 0.199 & 0.219 \\
 -it & NoExplain* & 2.7 & 4.1 & \textbf{\underline{8.8}} & 0.278 & 0.335 & 0.468 & 0.328 & \textbf{0.341} \\
 & CoT1 & 7.6 & 8.3 & 5.8 & 0.294 & \textbf{0.371} & 0.348 & 0.422 & 0.332 \\
 & CoT2 & 7.4 & 7.9 & 5.8 & \textbf{0.298} & \textbf{0.371} & 0.353 & 0.414 & 0.335 \\ [0.5em]
 DeepSeek- & Mastropaolo-T2 & 1.3 & 2.5 & 3.7 & 0.141 & 0.163 & 0.238 & 0.185 & 0.174 \\
 Coder-V2- & NoExplain & 4.0 & 3.7 & \textbf{6.2} & 0.263 & 0.327 & 0.390 & 0.338 & 0.313 \\
 Lite-Instruct & CoT1* & 7.6 & 4.4 & 5.7 & \textbf{0.287} & \textbf{0.356} & 0.362 & 0.394 & \textbf{0.327} \\
 & CoT2 & 9.4 & 4.6 & 5.4 & 0.282 & 0.354 & 0.329 & 0.404 & 0.314 \\ [0.5em]
 Llama-3.1-70B & Mastropaolo-T2 & 5.4 & 5.1 & 7.1 & 0.279 & 0.335 & 0.393 & 0.369 & 0.326 \\
 -Instruct & NoExplain* & 5.6 & 4.5 & 7.4 & \textbf{0.308} & \textbf{\underline{0.382}} & 0.389 & 0.408 & \textbf{0.352} \\
 & CoT1 & 8.2 & 7.6 & \textbf{7.6} & 0.307 & 0.380 & 0.355 & 0.430 & 0.341 \\
 & CoT2 & 8.3 & 7.5 & 7.2 & 0.302 & 0.376 & 0.348 & 0.434 & 0.338 \\ [0.5em]
 GPT-4o-mini & Mastropaolo-T2 & 3.2 & 2.5 & 2.9 & 0.191 & 0.229 & 0.324 & 0.253 & 0.235 \\
 -2024-07-18 & NoExplain* & 4.5 & 3.8 & \textbf{8.7} & \textbf{\underline{0.311}} & \textbf{0.380} & 0.438 & 0.386 & \textbf{\underline{0.365}} \\
 & CoT1 & 9.4 & 4.5 & 7.4 & 0.294 & 0.366 & 0.344 & 0.437 & 0.335 \\
 & CoT2 & 10.1 & 4.7 & 7.2 & 0.293 & 0.363 & 0.331 & 0.439 & 0.330 \\
  \bottomrule
\end{tabular}
\end{table}

\begin{table}
  \caption{Correlation between Exact Match and other metrics}
  \label{tab:correlation-with-EM}
  \begin{tabular}{llllll}
    \toprule
            & BLEU      & CrystalBLEU & BLEU & CrystalBLEU & LEMOD \\
            & WholeCode & WholeCode   & diff & diff        & LineF1 \\
    \midrule
    Python dataset & 0.007 & 0.076 & 0.827 & 0.788 & \textbf{0.837} \\
    Java dataset   & 0.004 & 0.081 & 0.693 & 0.653 & \textbf{0.748} \\
  \bottomrule
\end{tabular}
\end{table}

A possible question on this topic might be if EM is the most intuitive metric, why do we need the other metrics? To answer this, let’s examine Table~\ref{tab:rq2-python-results} more closely. As noted earlier, we created two slightly different prompts, CoT1 and CoT2, to test how a small change in the prompt might impact our evaluation. Ideally, a robust evaluation metric should report similar values for both prompts. However, as shown in Table~\ref{tab:rq2-python-results}, for Llama-3.1-8B and DeepSeek-Coder, there is a significant discrepancy between the values reported by Exact Match. In contrast, the other metrics remain more consistent, demonstrating their greater robustness compared to EM. Furthermore, because Exact Match does not account for the similarity between the two code outputs, it is not a suitable metric for assessing SATD items that are difficult to address. In such cases, EM is inadequate for determining which model has a better performance in SATD repayment. To conclude, while EM is a reliable measure for evaluating SATD repayment, it is not as robust as the other three metrics and cannot provide a similarity score for cases where the models fail to produce a fully matched answer.

In addition to the metrics discussed, Tables \ref{tab:rq2-python-results} and \ref{tab:rq2-java-results} also show the average number of deleted and inserted lines for each method. As expected, when using chain-of-thought prompts (CoT1 and CoT2), all models tend to make more changes to the code, either by deleting or inserting/updating lines. However, this increased code manipulation does not lead to a consistent improvement across all models and metrics when using the chain-of-thought approach. In fact, for both datasets, the highest score for each metric across all models and prompts (underlined values in Tables~\ref{tab:rq2-python-results} and \ref{tab:rq2-java-results}) is achieved by the NoExplain prompt, with the exception of Exact Match in the Python dataset, where Mastropaolo-T2 achieves the highest value.

Our LEMOD metric reports three values for each method: LineP, LineR, and LineF1. In both datasets, for each model, the NoExplain prompt achieves the highest LineP, while the two CoT prompts yield the highest LineR. The only exception is in the Python dataset when we use DeepSeek-Coder, where CoT1 achieves a slightly higher LineP than the NoExplain prompt.

Since the Mastropaolo-T2 template achieves the lowest values for most metrics and models (38 out of 40), incorporating ``\#\#\#'' separators in prompts is a good practice to help models better distinguish between different parts of a prompt. Among the five models studied, Llama-3.1-70B-Instruct is the most robust, meaning that the choice of prompt approach has the least effect on its results compared to the other models.

As explained earlier, we use the results from this research question to select the best prompt for each model, which will be used in RQ3 and RQ4. The selected prompt for each model is marked with an asterisk (*).

\begin{tcolorbox}[enhanced,width=6in,size=fbox,drop shadow southwest,sharp corners]

\textit{RQ2 Summary: While applying BLEU or CrystalBLEU to the entire code for evaluating generated code in SATD repayment results in unpredictable values, our proposed modified versions, BLEU-diff and CrystalBLEU-diff, yield significantly more accurate and explainable results. Additionally, our newly proposed evaluation metric, LEMOD, offers even greater detail while maintaining strong alignment with EM, BLEU-diff, and CrystalBLEU-diff.}

\end{tcolorbox}


\subsection{RQ3 \rqthree}\label{subsec:results-rq3}

Table~\ref{tab:mastropaolo-dataset} summarizes the number of instances in the Mastropaolo dataset before and after applying the two new filters introduced in RQ1. Approximately 60\% of the items are removed through this filtering process. Table~\ref{tab:rq3-results} compares the performance of two approaches on the original and filtered Mastropaolo datasets.

In the first approach, GPT-4o-mini paired with the NoExplain prompt achieves significantly higher results across all four metrics on the filtered dataset compared to the original dataset. In the second approach, fine-tuning the CodeT5p-770m model on the filtered training data yields consistently better performance across all four metrics than fine-tuning on the controlled set, which contains 1,391 randomly sampled instances from the original training data. These findings highlight the effectiveness of filtering SATD repayment datasets in improving performance for both prompt-based LLMs and those undergoing continuous training, i.e., fine-tuning.

\begin{table}
  \caption{Number of instances in the Mastropaolo dataset before and after our new filters}
  \label{tab:mastropaolo-dataset}
  \begin{tabular}{lcc}
    \toprule
    Part & Original dataset & Filtered dataset \\
    \midrule
    Train       & 3,537 & 1,391 \\
    Evaluation  & 502  & 210  \\
    Test        & 1,000 & 386  \\
  \bottomrule
\end{tabular}
\end{table}

\begin{table}[t]
  \caption{Performance comparison of models using the original and filtered Mastropaolo dataset}
  \label{tab:rq3-results}
  \resizebox{15cm}{!}{
  \begin{tabular}{lllllll}
    \toprule
    Model & Approach & Test    & Exact & BLEU & CrystalBLEU & LEMOD \\
          &          & dataset & Match & diff & diff        & LineF1 \\
    \midrule
    GPT-4o-mini & NoExplain prompt template & original Mastropaolo & 4.6 & 0.228 & 0.289 & 0.231 \\
    GPT-4o-mini & NoExplain prompt template & filtered Mastropaolo & \textbf{7.5} & \textbf{0.276} & \textbf{0.344} & \textbf{0.266} \\ [0.5em]
    CodeT5p-770m & Fine-tune on controlled Mastropaolo & filtered Mastropaolo & 4.7 & 0.194 & 0.247 & 0.220 \\
    CodeT5p-770m & Fine-tune on filtered Mastropaolo & filtered Mastropaolo & \textbf{6.2} & \textbf{0.201} & \textbf{0.250} & \textbf{0.226} \\
  \bottomrule
\end{tabular}
}
\end{table}

\begin{tcolorbox}[enhanced,width=6in,size=fbox,drop shadow southwest,sharp corners]

\textit{RQ3 Summary: Applying our proposed filters to the Mastropaolo dataset results in a cleaner dataset, which enhances performance in LLM-based automated SATD repayment, whether through prompt engineering or fine-tuning a model.} 

\end{tcolorbox}


\subsection{RQ4 \rqfour}\label{subsec:results-rq4}

Tables \ref{tab:rq4-python-results} and \ref{tab:rq4-java-results} summarize the performance of the five prompt-based methods and the two fine-tuned models on the Python and Java test sets. 

\begin{table}[h]
  \caption{Performance of the five prompt-based methods and the two fine-tuned models on the Python test set (5,745 items)}
  \label{tab:rq4-python-results}
  \begin{tabular}{llllllll}
    \toprule
    Model & Prompt   &  Exact & BLEU & CrystalBLEU & \multicolumn{3}{c}{LEMOD} \\
    \cline{6-8}
          & Template &  Match & diff & diff        & LineP & LineR & LineF1 \\
    \midrule
Llama-3.1-8B-Instruct & CoT1 & 8.2 & 0.333 & 0.403 & 0.380 & 0.355 & 0.327 \\
Gemma-2-9B-it & NoExplain & \textbf{10.1} & 0.326 & 0.381 & 0.419 & 0.339 & 0.336 \\
DeepSeek-Coder-V2-Lite-Instruct & CoT1 & 7.2 & 0.291 & 0.360 & 0.328 & 0.328 & 0.288 \\
Llama-3.1-70B-Instruct & NoExplain & 8.9 & \textbf{0.376} & \textbf{0.448} & 0.419 & 0.391 & \textbf{0.368} \\
GPT-4o-mini-2024-07-18 & CoT1 & 9.4 & 0.356 & 0.431 & 0.382 & 0.382 & 0.344 \\ [0.5em]
Fine-tuned CodeT5p-220m-py & N/A & 9.2 & 0.214 & 0.248 & 0.307 & 0.233 & 0.236 \\
Fine-tuned CodeT5p-770m-py & N/A & 9.7 & 0.255 & 0.295 & 0.356 & 0.271 & 0.275 \\
  \bottomrule
\end{tabular}
\end{table}

\begin{table}[h]
  \caption{Performance of the five prompt-based methods and the two fine-tuned models on the Java test set (9,013 items)}
  \label{tab:rq4-java-results}
  \begin{tabular}{llllllll}
    \toprule
    Model & Prompt   &  Exact & BLEU & CrystalBLEU & \multicolumn{3}{c}{LEMOD} \\
    \cline{6-8}
          & Template &  Match & diff & diff        & LineP & LineR & LineF1 \\
    \midrule
Llama-3.1-8B-Instruct & NoExplain & 5.6 & 0.282 & 0.345 & 0.402 & 0.342 & 0.326 \\
Gemma-2-9B-it & NoExplain & \textbf{8.1} & 0.274 & 0.325 & 0.463 & 0.325 & 0.335 \\
DeepSeek-Coder-V2-Lite-Instruct & CoT1 & 4.7 & 0.267 & 0.334 & 0.338 & 0.374 & 0.306 \\
Llama-3.1-70B-Instruct & NoExplain & 5.9 & 0.308 & \textbf{0.379} & 0.384 & 0.404 & 0.346 \\
GPT-4o-mini-2024-07-18 & NoExplain & 7.3 & \textbf{0.309} & 0.376 & 0.417 & 0.374 & \textbf{0.351} \\ [0.5em]
Fine-tuned CodeT5p-220m & N/A & 6.7 & 0.227 & 0.284 & 0.395 & 0.287 & 0.287 \\
Fine-tuned CodeT5p-770m & N/A & 7.6 & 0.249 & 0.309 & 0.411 & 0.305 & 0.305 \\
  \bottomrule
\end{tabular}
\end{table}

The results show that Gemma-2-9B consistently achieves the highest Exact Match (EM) scores across both datasets, with 10.1\% on the Python data and 8.1\% on the Java data. These scores are significantly higher than the 1.19\% reported by Mastropaolo et al. for the best-performing prompt-based model. As for the fine-grained evaluation, for the Python dataset, Llama-3.1-70B demonstrates the best performance across BLEU-diff, CrystalBLEU-diff, and LineF1, reflecting its superior ability to capture nuanced changes and maintain line-level accuracy. For the Java dataset, GPT-4o-mini and Llama-3.1-70B perform comparably, outperforming other models in BLEU-diff, CrystalBLEU-diff, and LineF1.

For the fine-tuned models, leveraging a large training dataset allows smaller models to achieve EM scores comparable to those of prompt-based models. The best-performing fine-tuned model, Fine-tuned CodeT5p-770M(-py), achieves an EM of 9.7\% on the Python test set and 7.6\% on the Java test set, significantly surpassing the 2.3\% EM reported by Mastropaolo et al. for their best fine-tuned model. However, for the other three metrics, i.e., BLEU-diff, CrystalBLEU-diff, and LineF1, a noticeable performance gap remains. This gap is somewhat reduced when leveraging the 770M base model, but it is still significant.

In the above results, we do not differentiate SATD items based on their difficulty level. However, LLM performance may vary significantly between easy-to-address and hard-to-address SATD items. Understanding how different LLMs perform on these two groups can offer deeper insights into their strengths and limitations. To explore this, we conduct a follow-up analysis comparing the performance of LLMs on both groups. The datasets are categorized into easy and hard groups based on the following definitions:

\begin{itemize}
    \item \textbf{Easy-to-address SATD:} An item where, in the diff between the input code and the ground truth, at most two lines are inserted.
    \item \textbf{Hard-to-address SATD:} An item where, in the diff between the input code and the ground truth, at least three lines are inserted.
\end{itemize}

The number of deleted lines is not considered, as deleting lines is generally less challenging than inserting new ones. Based on these criteria, 43.3\% of the Python dataset and 48.6\% of the Java dataset are classified as easy-to-address SATD items. Tables \ref{tab:rq4-python-results-on-each-group} and \ref{tab:rq4-java-results-on-each-group} present the scores on easy and hard groups, along with those on the whole test set.

\begin{table}[h]
  \caption{Performance of models on each subset of the Python test set (easy: 43.3\%, hard: 56.7\%)}
  \label{tab:rq4-python-results-on-each-group}
  \begin{tabular}{lllllllll}
    \toprule
    Model & Prompt   & Subset & Exact & BLEU & CrystalBLEU & \multicolumn{3}{c}{LEMOD} \\
    \cline{7-9}
          & Template &       & Match & diff & diff        & LineP & LineR & LineF1 \\
    \midrule
     &  & easy & 7.5 & 0.459 & 0.519 & 0.454 & 0.516 & 0.447 \\
    Llama3.1-8B & CoT1 & hard & 0.7 & 0.236 & 0.315 & 0.324 & 0.233 & 0.235 \\
     &  & all & 8.2 & 0.333 & 0.403 & 0.380 & 0.355 & 0.327 \\ [0.4em]
     &  & easy & \textbf{9.1} & 0.480 & 0.537 & 0.509 & 0.515 & 0.482 \\
    Gemma-2-9B & NoExplain & hard & 1.0 & 0.208 & 0.262 & 0.350 & 0.205 & 0.225 \\
     &  & all & 10.1 & 0.326 & 0.381 & 0.419 & 0.339 & 0.336 \\ [0.4em]
     &  & easy & 6.5 & 0.383 & 0.438 & 0.388 & 0.466 & 0.386 \\
    DeepSeek-Coder-V2-Lite-Inst. & CoT1 & hard & 0.7 & 0.220 & 0.300 & 0.281 & 0.222 & 0.213 \\
     &  & all & 7.2 & 0.291 & 0.360 & 0.328 & 0.328 & 0.288 \\ [0.4em]
     &  & easy & 8.1 & \textbf{0.512} & \textbf{0.575} & 0.497 & 0.572 & \textbf{0.502} \\
    Llama3.1-70B & NoExplain & hard & 0.8 & \textbf{0.271} & 0.352 & 0.359 & 0.253 & \textbf{0.265} \\
     &  & all & 8.9 & 0.376 & 0.448 & 0.419 & 0.391 & 0.368 \\ [0.4em]
     &  & easy & 8.5 & 0.471 & 0.526 & 0.463 & 0.550 & 0.466 \\
    gpt-4o-mini-2024-07-18 & CoT1 & hard & 0.9 & 0.269 & \textbf{0.358} & 0.321 & 0.254 & 0.250 \\
     &  & all & 9.4 & 0.356 & 0.431 & 0.382 & 0.382 & 0.344 \\ [0.4em]
     &  & easy & 8.4 & 0.357 & 0.393 & 0.410 & 0.388 & 0.375 \\
    Fine-tuned CodeT5p-220m-py & N/A & hard & 0.8 & 0.105 & 0.137 & 0.229 & 0.115 & 0.130 \\
     &  & all & 9.2 & 0.214 & 0.248 & 0.307 & 0.233 & 0.236 \\ [0.4em]
     &  & easy & 9.0 & 0.416 & 0.459 & 0.452 & 0.443 & 0.424 \\
    Fine-tuned CodeT5p-770m-py & N/A & hard & 0.7 & 0.131 & 0.170 & 0.283 & 0.140 & 0.161 \\
     &  & all & 9.7 & 0.255 & 0.295 & 0.356 & 0.271 & 0.275 \\
  \bottomrule
\end{tabular}
\end{table}

\begin{table}[h]
  \caption{Performance of models on each subset of the Java test set (easy: 48.6\%, hard: 51.4\%)}
  \label{tab:rq4-java-results-on-each-group}
  \begin{tabular}{lllllllll}
    \toprule
    Model & Prompt   & Subset & Exact & BLEU & CrystalBLEU & \multicolumn{3}{c}{LEMOD} \\
    \cline{7-9}
          & Template &       & Match & diff & diff        & LineP & LineR & LineF1 \\
    \midrule
     &  & easy & 5.1 & 0.372 & 0.445 & 0.427 & 0.468 & 0.411 \\
    Llama3.1-8B & NoExplain & hard & 0.5 & 0.196 & 0.249 & 0.379 & 0.223 & 0.245 \\
     &  & all & 5.6 & 0.282 & 0.345 & 0.402 & 0.342 & 0.326 \\ [0.4em]
     &  & easy & \textbf{7.4} & 0.398 & 0.463 & 0.509 & 0.471 & \textbf{0.454} \\
    Gemma-2-9B & NoExplain & hard & 0.6 & 0.156 & 0.195 & 0.420 & 0.186 & 0.221 \\
     &  & all & 8.1 & 0.274 & 0.325 & 0.463 & 0.325 & 0.335 \\ [0.4em]
     &  & easy & 4.3 & 0.313 & 0.377 & 0.348 & 0.482 & 0.359 \\
    DeepSeek-Coder-V2-Lite-Inst. & CoT1 & hard & 0.4 & 0.224 & 0.293 & 0.329 & 0.272 & 0.256 \\
     &  & all & 4.7 & 0.267 & 0.334 & 0.338 & 0.374 & 0.306 \\ [0.4em]
     &  & easy & 5.4 & 0.378 & 0.449 & 0.402 & 0.534 & 0.417 \\
    Llama3.1-70B & NoExplain & hard & 0.5 & \textbf{0.241} & \textbf{0.312} & 0.366 & 0.281 & \textbf{0.279} \\
     &  & all & 5.9 & 0.308 & 0.379 & 0.384 & 0.404 & 0.346 \\ [0.4em]
     &  & easy & 6.5 & \textbf{0.400} & \textbf{0.472} & 0.452 & 0.502 & 0.438 \\
    gpt-4o-mini-2024-07-18 & NoExplain & hard & 0.8 & 0.224 & 0.286 & 0.383 & 0.253 & 0.268 \\
     &  & all & 7.3 & 0.309 & 0.376 & 0.417 & 0.374 & 0.351 \\ [0.4em]
     &  & easy & 6.2 & 0.321 & 0.395 & 0.420 & 0.413 & 0.382 \\
    Fine-tuned CodeT5p-220m & N/A & hard & 0.5 & 0.137 & 0.179 & 0.370 & 0.167 & 0.196 \\
     &  & all & 6.7 & 0.227 & 0.284 & 0.395 & 0.287 & 0.287 \\ [0.4em]
     &  & easy & 7.0 & 0.351 & 0.427 & 0.443 & 0.442 & 0.410 \\
    Fine-tuned CodeT5p-770m & N/A & hard & 0.6 & 0.152 & 0.197 & 0.379 & 0.176 & 0.205 \\
     &  & all & 7.6 & 0.249 & 0.309 & 0.411 & 0.305 & 0.305 \\
  \bottomrule
\end{tabular}
\end{table}

As expected, across all approaches considered, the majority of the EM score (over 90\%) originates from the easy-to-address group, even though more than half of the SATD items are classified as hard-to-address. However, for the other three metrics, all models achieve scores from both groups, although their scores in the easy group are higher than in the hard group. In other words, while EM predominantly reflects performance on the easy group, the other three metrics are influenced by both groups. Consequently, using EM as a performance metric for the hard group is not recommended, as the scores are typically very low (often below 1\%), making them highly sensitive to minor variations. For instance, correctly addressing just a few additional items by chance can disproportionately influence model rankings. A clear example of this issue is observed with CodeT5p-770m-py, which, despite being over three times larger than CodeT5p-220m-py, achieves a lower EM score on the hard group in the Python dataset—an outcome that defies logical expectations. The alternative fine-grained metrics provide a more reliable and rational distinction in performance between these two models.

According to the above analysis, we use all four metrics to measure effectiveness on the easy group, and the three fine-grained metrics for the hard group. The best performing model for each group and metric combination is highlighted in Tables~\ref{tab:rq4-python-results-on-each-group} and \ref{tab:rq4-java-results-on-each-group}. On the Python dataset’s easy group, Llama3.1-70B with the NoExplain prompt achieves the highest scores in three of the four metrics, while Gemma-2-9B attains the best EM score. Notably, Gemma-2-9B surpasses GPT4o-mini across all four metrics in this easy group. On the hard group of the Python dataset, Llama3.1-70B with the NoExplain prompt performs best in CrystalBLEU-diff and LineF1, while GPT4o-mini leads in BLEU-diff. For the Java dataset’s easy group, Gemma-2-9B ranks first in EM and LineF1, second in BLEU-diff and CrystalBLEU-diff, while GPT4o-mini leads in the latter two metrics. Finally, in the Java dataset’s hard group, Llama3.1-70B outperforms all other models on all three evaluated metrics. The fine-tuned models, especially CodeT5p-770m, while providing comparable performance to larger prompt-based models on the easy groups, fall behind by a large margin on the hard groups.

\begin{tcolorbox}[enhanced,width=6in,size=fbox,drop shadow southwest,sharp corners]

\textit{RQ4 Summary: While fine-tuned smaller models achieve the same level of Exact Match as larger models (with prompt engineering), they lag behind in the other three fine-grained evaluation metrics, highlighting their limitations in addressing more complex SATDs. Also, the EM score is not recommended for evaluating models in hard-to-address SATDs. The fine-grained evaluation techniques provide more reliable results.} 

\end{tcolorbox}

\section{Discussion}\label{sec:discussion}
\subsection{Optimal Performance with an Oracle Template}\label{subsec:oracle-prompt}
Prompt engineering is widely recognized as a critical factor in the performance of LLMs~\cite{chen2024unleashing,Sahoo2024}. As demonstrated in the results of RQ2 (ref. Tables~\ref{tab:rq2-python-results} and \ref{tab:rq2-java-results}), the four prompt templates lead to varying levels of performance, with different prompts excelling under different settings. This variability raises an intriguing question: if the best prompt could always be selected for each scenario, what would the performance of the models be? To explore this potential upper bound enabled by a mix of prompts, we evaluated the hypothetical scenario of using an \textit{oracle template}, i.e., a method that dynamically selects the optimal prompt template for each SATD item. 

Table~\ref{tab:oracle-performance} compares the Exact Match scores of models using the oracle template against those achieved with individual prompt templates. Across all models, the oracle template consistently delivers superior performance, surpassing the best single prompt template. Remarkably, models like Llama-3.1-8B and DeepSeek-Coder, which otherwise exhibit the lowest performance, outperform high-performing models such as Llama-3.1-70B and GPT-4o-mini when using the oracle template. The results indicate that while the single best-performing template addresses most SATD items handled by the others, certain SATD items are only captured by one or more of the alternative templates. This suggests that \textbf{developing a dynamic model capable of selecting the most effective prompt template for each SATD item could unlock significant improvements in the performance of LLMs for SATD repayment.} Exploring methods to achieve this remains a promising direction for future research.

\begin{table}
  \caption{Performance comparison of oracle template vs. individual prompt templates across models}
  \label{tab:oracle-performance}
  \begin{tabular}{llll}
    \toprule
    Model & Prompt Template & EM in Python & EM in Java \\
    \midrule
    Llama-3.1-8B-Instruct & Mastropaolo-T2 & 5.1 & 3.6 \\
     & NoExplain & 7.7 & 5.7 \\
     & CoT1 & 9.1 & 5.4 \\
     & CoT2 & 7.1 & 5.3 \\
     & \textbf{Oracle} & \textbf{12.1} & \textbf{10.0} \\ [0.5em]
    Gemma-2-9B-it & Mastropaolo-T2 & 7.3 & 5.8 \\
     & NoExplain & 9.9 & 8.8 \\
     & CoT1 & 6.4 & 5.8 \\
     & CoT2 & 6.8 & 5.8 \\
     & \textbf{Oracle} & \textbf{14.0} & \textbf{10.6} \\ [0.5em]
    DeepSeek-Coder-V2-Lite-Instruct & Mastropaolo-T2 & 3.8 & 3.7 \\
     & NoExplain & 5.6 & 6.2 \\
     & CoT1 & 7.7 & 5.7 \\
     & CoT2 & 5.8 & 5.4 \\
     & \textbf{Oracle} & \textbf{11.3} & \textbf{9.3} \\ [0.5em]
    Llama-3.1-70B-Instruct & Mastropaolo-T2 & 10.5 & 7.1 \\
     & NoExplain & 10.0 & 7.4 \\
     & CoT1 & 10.0 & 7.6 \\
     & CoT2 & 9.7 & 7.2 \\
     & \textbf{Oracle} & \textbf{14.3} & \textbf{11.1} \\ [0.5em]
    GPT-4o-mini-2024-07-18 & Mastropaolo-T2 & 5.1 & 2.9 \\
     & NoExplain & 9.2 & 8.7 \\
     & CoT1 & 9.6 & 7.4 \\
     & CoT2 & 9.3 & 7.2 \\
     & \textbf{Oracle} & \textbf{12.9} & \textbf{10.6} \\
  \bottomrule
\end{tabular}
\end{table}


\subsection{How Well do Models Cover SATD Repayments?}\label{subsec:repayment-coverage}
In RQ2, we conducted 20 experiments on each dataset, using five models and four prompt templates. This section analyzes the results to assess SATD repayment coverage under three settings: (1) \textbf{Setting \#1:} SATD repayments are addressed by all 20 model-template combinations. This examines the sensitivity of SATD repayment success to varying combinations of LLMs and prompts. (2) \textbf{Setting \#2:} SATD repayments are addressed by at least one model-template combination. This represents an upper bound for SATD repayment if we could always select the best-performing combination. (Note: This differs from the results in Section~\ref{subsec:oracle-prompt}, as it allows the model to vary, whereas the prior section fixes the LLM.) (3) \textbf{Setting \#3:} SATD repayments are addressed by only one model-template combination. This identifies non-overlapping cases, where only highly specialized model-template combinations can address certain items. Here, addressed refers to achieving an EM score of 1 for a given SATD item. The results are summarized in Table~\ref{tab:repayment-coverage}. 

Interestingly, only three items in the Python dataset and one item in the Java dataset are addressed by all model-template combinations (Setting \#1). This suggests that \textbf{SATD repayment success is highly sensitive to the choice of LLM and prompt}. Encouragingly, 195 items (19.5\%) in Python and 150 items (15\%) in Java are addressed by at least one model-template combination (Setting \#2), indicating \textbf{promising progress of prompt-based methods toward practical automated SATD repayment}. Finally, 27 items in Python and 20 in Java are addressed by only one specific model-template combination (Setting \#3), highlighting that \textbf{certain SATD items require highly specialized approaches (specific prompts/models) for effective handling}.

\begin{table}[h]
  \caption{SATD repayment coverage in RQ2 (out of 1000 samples)}\label{tab:repayment-coverage}
  \begin{tabular}{lll}
    \toprule
    SATD items that addressed by & Python & Java \\
    \midrule
    All 20 model-templates & 3 & 1 \\
    At least one model-template & 195 & 150 \\
    At most one model-template & 27 & 20 \\
  \bottomrule
\end{tabular}
\end{table}


\subsection{The Chosen of Evaluation Metric(s)}\label{subsec:which-metric}
In this study, we utilize four metrics to evaluate automated SATD repayment: EM, BLEU-diff, CrystalBLEU-diff, and LEMOD. A key question is which metric is most suitable for practical use. To answer this question, we analyze the correlations among the four metrics using the results from RQ2, as summarized in Table~\ref{tab:evaluation-metrics-correlation}.

\begin{table} [h]
  \caption{The correlation between four evaluation metrics from the results of RQ2 for Python and Java datasets separately}
  \label{tab:evaluation-metrics-correlation}
  \begin{tabular}{lr|cccc}
    \toprule
    & & & \multicolumn{3}{c}{Python dataset} \\
    & & EM & BLEU-diff & CrystalBLEU-diff & LineF1 \\
    \midrule
     & EM & - & 0.827 & 0.788 & 0.837 \\
     & BLEU-diff & 0.693 & - & 0.997 & 0.996 \\
    Java & CrystalBLEU-diff & 0.653 & 0.997 & - & 0.989 \\
    dataset & LineF1 & 0.748 & 0.987 & 0.976 & - \\
  \bottomrule
\end{tabular}
\end{table}

As shown in Table~\ref{tab:evaluation-metrics-correlation}, BLEU-diff, CrystalBLEU-diff, and LEMOD-LineF1 exhibit very high correlations (above 0.97) in both datasets. This suggests that the three metrics produce similar evaluations. However, LEMOD provides additional insights by offering line-level precision (LineP) and recall (LineR) and is generally easier to interpret than BLEU-diff and CrystalBLEU-diff, making it a more comprehensive choice.
In contrast, the EM metric shows moderate correlation (0.65–0.84) with the other three metrics, as it fundamentally differs in focus. EM measures the proportion of fully addressed items, which is particularly valuable in scenarios where the generated code must be directly usable without further refinement. This distinction makes EM a critical complement to the other metrics, especially for applications emphasizing correctness over partial matches or incremental improvements.
Therefore, \textbf{we recommend using both EM and LEMOD as a complementary set of metrics for future evaluations of automated SATD repayment.} Together, they balance the strictness of exact matches with the flexibility and detailed insights offered by line-level evaluations.
Lastly, preprocessing plays a vital role in ensuring the accuracy of metric calculations. In this study, we removed ICD elements (Imports, Comments, and Docstrings/JavaDoc) for all metrics. Additionally, for EM, we applied specific formatting adjustments, including removing all whitespace for Java and standardizing formatting for Python.

\section{Threats to Validity}\label{sec:threats}
\noindent \textbf{Internal Validity.}
In the final step of our filtering pipeline to create a clean dataset, we used the Llama-3-70B-Instruct model, which achieved a precision of 0.830 for Python and 0.926 for Java. This indicates that the dataset is not completely clean, as some items with incorrect ground truth remain, potentially negatively affecting the reported performance. Additionally, due to the model's less-than-perfect recall, we have missed including some SATDs in our filtered datasets. These overlooked items are often difficult for models to address, so their exclusion may have led to an overestimation of our model's performance.  As these two threats counteract each other, they alleviate each other’s impact on our reported scores.

\noindent \textbf{External Validity.}
In this study, we focused on within-method SATD repayment. Consequently, the reported performance for automated SATD repayment may not generalize to class-level or file-level SATD, as these types can impact more lines of code and could be more challenging to address. Moreover, our results pertain to SATD in Python and Java, so they may not generalize to other programming languages. Another threat to external validity is that some of the filtering steps employed in RQ1 (such as removing SATDs with non-ASCII characters and those with two or fewer words) reduce the representativeness of our SATD repayment dataset. However, as explained in Section~\ref{subsec:RQ1}, these steps were necessary.

\noindent \textbf{Construct Validity.}
A common threat to the validity of leveraging recent LLMs, such as Llama, Gemma, and GPT-4o, is the possibility that they may have already encountered the target code during their pre-training phase, raising concerns about data leakage. We believe that, although the target code may have appeared in the pre-training data, it is less likely to have appeared alongside its corresponding input code (that includes the SATD comment) either consecutively in the pre-training data or as part of a question-answer pair in the instruction fine-tuning data. To the best of our knowledge, our SATD repayment dataset is the first of its kind for the Python programming language, and our Java SATD repayment dataset, containing 97,347 items, introduces many new items, even if those LLMs were already trained on the Mastropaolo dataset, which contains only 5,039 items.

\section{Conclusion and Future Work}\label{sec:conclusion}
This work aims to benchmark and understand the effectiveness of LLMs in automating SATD repayment, i.e., addressing SATDs. To achieve this goal, we identified and resolved three critical challenges in creating benchmarks and evaluation methods for automated SATD repayment. Our contributions include a novel pipeline for collecting and cleaning SATD repayments, resulting in the largest and most robust datasets for Python and Java. Additionally, we introduced three new evaluation metrics to address the limitations of existing methods. Leveraging these benchmarks and metrics, we evaluated the performance of fine-tuning and prompt-based LLMs on this task and demonstrated the positive impact of cleaner datasets for both approaches.

Our experiments revealed that fine-tuning smaller models with hundreds of millions of parameters lags behind prompt-based approaches leveraging larger models with billions of parameters. However, fine-tuning outcomes could be enhanced by employing larger models similar to those used in prompt-based setups. In terms of exact match (EM) score, the highest achieved by current LLMs is 8.1\% on Java SATDs and 10.1\% on Python SATDs.

We further explored the potential for improving LLM-based approaches by learning optimal model-prompt combinations and provided guidance for practitioners on selecting the most suitable evaluation metrics. We believe that our new data preprocessing pipeline, benchmark datasets, and evaluation metrics will enable future research to develop more effective and scalable solutions for automated SATD repayment.

\begin{acks}
We acknowledge the support of the Natural Sciences and Engineering Research Council of Canada (NSERC), [funding reference number: RGPIN-2019-05071].
\end{acks}

\bibliographystyle{ACM-Reference-Format}
\bibliography{_main}


\end{document}